\documentclass[%
 reprint,
 superscriptaddress,
 longbibliography,
 amsmath,amssymb,
 prx
]{revtex4-2}

\usepackage{graphicx}
\usepackage{dcolumn}
\usepackage{bm}

\usepackage{siunitx}
\usepackage{braket}
\usepackage{hyperref}
\DeclareSIUnit{\dBm}{dBm}
\begin{document}

\preprint{APS/123-QED}

\title{Microwave Control of the Tin-Vacancy Spin Qubit in Diamond with a Superconducting Waveguide}

\newcommand{\kitphi}{Physikalisches Institut (PHI), Karlsruhe Institute of Technology (KIT), Wolfgang-Gaede-Straße 1, 76131 Karlsruhe, Germany}

\newcommand{\saarb}{Department of Physics, Saarland University, Campus E2 6, 66123 Saarbrücken, Germany}

\newcommand{\leipzig}{Division of Applied Quantum Systems, Felix-Bloch-Institute for Solid State Physics, University of Leipzig, 04103 Leipzig, Germany}

\newcommand{\kassel}{Institute of Nanostructure Technologies and Analytics (INA), Center for Interdisciplinary Nanostructure Science and Technology (CINSaT), University of Kassel, Heinrich-Plett-Straße 40, 34132 Kassel, Germany}

\newcommand{\kitiqmt}{Institute for Quantum Materials and Technologies (IQMT), Karlsruhe Institute of Technology (KIT), Herrmann-von-Helmholtz Platz 1, 76344 Eggenstein-Leopoldshafen, Germany}

\author{Ioannis Karapatzakis}
\thanks{These two authors contributed equally}
\affiliation{\kitphi}%

\author{Jeremias Resch}
\thanks{These two authors contributed equally}
\affiliation{\kitphi}%

\author{Marcel Schrodin}
\affiliation{\kitphi}%

\author{Philipp Fuchs}
\affiliation{\saarb}%

\author{Michael Kieschnick}
\affiliation{\leipzig}%

\author{Julia Heupel}
\affiliation{\kassel}%

\author{Luis Kussi}
\affiliation{\kitphi}%

\author{Christoph Sürgers}
\affiliation{\kitphi}%

\author{Cyril Popov}
\affiliation{\kassel}%

\author{Jan Meijer}
\affiliation{\leipzig}%

\author{Christoph Becher}
\affiliation{\saarb}%

\author{Wolfgang Wernsdorfer}
\affiliation{\kitphi}
\affiliation{\kitiqmt}%

\author{David Hunger}
\email{david.hunger@kit.edu}
\affiliation{\kitphi}
\affiliation{\kitiqmt}%

\date{\today}

\begin{abstract}
Group-IV color centers in diamond are promising candidates for quantum networks due to their dominant zero-phonon line and symmetry-protected optical transitions that connect to coherent spin levels. The negatively charged tin-vacancy (SnV) center possesses long electron spin lifetimes due to its large spin-orbit splitting. However, the magnetic dipole transitions required for microwave spin control are suppressed, and strain is necessary to enable these transitions. Recent work has shown spin control of strained emitters using microwave lines that suffer from Ohmic losses, restricting coherence through heating. We utilize a superconducting coplanar waveguide to measure SnV centers subjected to strain, observing substantial improvement. A detailed analysis of the SnV center electron spin Hamiltonian based on the angle-dependent splitting of the ground and excited states is performed. We demonstrate coherent spin manipulation and obtain a Hahn echo coherence time of up to $T_2 = \SI{430}{\micro\second}$. With dynamical decoupling, we can prolong coherence to $T_2 = \SI{10}{\milli\second}$, about six-fold improved compared to earlier works. We also observe a nearby coupling $^{13}\mathrm{C}$ spin which may serve as a quantum memory. This substantiates the potential of SnV centers in diamond and demonstrates the benefit of superconducting microwave structures.
\end{abstract}

\maketitle


\section{\label{sec:sec1}Introduction}
Optically addressable coherent spins in the solid state are favorable candidates for the realization of quantum networks \cite{Kimble2008, Atatüre2018}. Among them, color centers in diamond have enabled demonstrations of the full set of functionalities in this context, including spin-photon \cite{Togan2010, Hensen2015} and spin-spin entanglement \cite{Bernien2013, knaut2023entanglement}, teleportation \cite{Hermans2022}, and the first demonstrations of a three-node network \cite{pompili2021}. Group-IV color centers in diamond have recently emerged as a promising system, in particular due to their excellent optical properties: Their inversion symmetry leads to a dominant zero-phonon line and reduced sensitivity against electrical noise. This has enabled e.g. the incorporation of silicon-vacancy centers in optical nanocavities \cite{Bhaskar2020, Stas2022} and the demonstration of quantum network elements \cite{Bhaskar2020, knaut2023entanglement, Bersin2024}. However, the presence of two orbital ground states opens a decoherence channel for the spin levels via phonon scattering. In this respect, the tin-vacancy (SnV) center offers advantages due to its large ground-state splitting of $>\SI{800}{\giga\hertz}$, such that phonons at this frequency can be frozen out at comparably high temperatures $>\SI{1}{\kelvin}$.
Reported dephasing times for all-optical control range from \SI{1.3}{\micro\second} \cite{Debroux2021} to \SI{5}{\micro\second} at \SI{2}{\kelvin} temperature \cite{Görlitz2022} with coherence times ranging from \SI{0.3}{\milli\second} using all-optical control \cite{Debroux2021} to \SI{1.6}{\milli\second} for microwave control \cite{Guo}, respectively.
Despite that, direct spin control by microwave fields is hampered by orthogonal orbital contributions to the Zeeman states. Recently, it has been shown that by introducing strain, orbital mixing can be induced and high-fidelity microwave control becomes possible \cite{Rosenthal,Guo}. This regime makes the level structure non-trivial and motivates an accurate mapping of the spin Hamiltonian, which is, however, still partially missing and treated inconsistently in literature. Especially the quenched Zeeman splitting of the quantized orbital momentum has not been precisely determined yet \cite{Rosenthal, Guo, Thiering_magneto_optic}.

In this paper, we characterize the negatively charged SnV center regarding its magneto-optical properties and determine the relevant components, such as the orbital quenching factors from a full fit of the electronic spin Hamiltonian. We explain the properties of the ground and excited state under strain and the relevant qubit transitions, as well as the influence of an external magnetic field. We demonstrate coherent control of the electron spin and reach a $T_2$ time of $\SI{10}{\milli\second}$ by using standard dynamical decoupling sequences. A simplified design and fabrication process demonstrates the potential of the SnV center for quantum computing, communication, and the establishment of robust quantum networks.

\begin{figure*}
\includegraphics{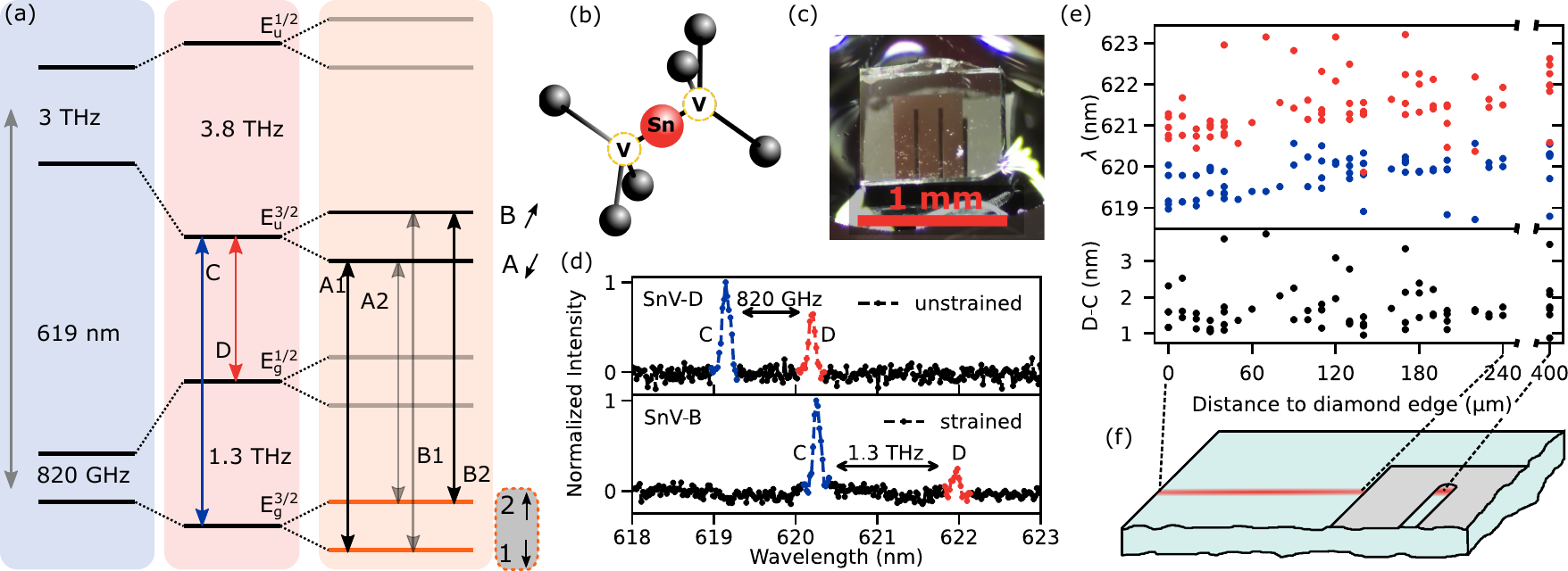}
\caption{\label{fig:fig_overwiev} A strained SnV center in a diamond membrane with a coplanar waveguide. (a) Energy level scheme of a strained SnV center in a magnetic field. The C and D transitions are depicted as blue and red arrows, respectively. A magnetic field further splits the spin states, and the black (grey) arrows indicate the allowed (forbidden) transitions. The qubit is indicated by orange-colored energy levels. (b) Schematic of the SnV center in diamond showing the interstitial Sn-atom neighbored by two vacancies inside the diamond lattice. (c) Optical image of the sample fixed onto a silicon substrate with an adhesive. The niobium coplanar waveguide is fabricated on top of the diamond for microwave control. (d) Photoluminescence spectra of an unstrained (SnV-D) and a strained (SnV-B) SnV center with the C (D) transition marked in blue (red). The measurements are performed at $\SI{4}{\kelvin}$ showing only the ground state splitting. (e) Top: Strain induced shift of the C and D transitions of SnV centers measured along the membrane from the edge of the diamond to center. Bottom: Corresponding ground state splitting $\mathrm{D}-\mathrm{C}$. The PL measurements are taken within the area marked in red on the diamond membrane schematic in (f), with the superconducting CPW depicted in grey.}
\end{figure*}

\section{\label{sec:sec2_sample_fab}Electronic structure of strained SnV centers}

It has been observed, that microwave control of Group-IV defects is only achievable under a strained environment of the emitters \cite{Guo,Rosenthal,Meesala_strain,Pingault_siv_strain, SukachevSiV}, explained by the orthogonality of the electron orbital states of the two spin qubit levels. However, recent work suggests that efficient microwave spin control is possible above the Larmor frequency of the free electron $\sim \SI{28}{\GHz\per\tesla}$ for specific combinations of the microwave $B_\mathrm{ac}$ and magnetic field $B_\mathrm{dc}$ orientations even in the absence of strain \cite{Pieplow_strain}. Still, $B_\mathrm{ac}$ is predefined by the design of the microwave antenna, and thus the main handle is the orientation of the $B_\mathrm{dc}$ field with respect to the defect's symmetry axis. The role of strain is to enhance the overall microwave response of the SnV electron spin by mixing the orthogonal orbitals and thus relaxing the magnetic dipole selection rules, allowing microwave control for almost all magnetic field orientations. Therefore, we intentionally induce strain into our sample.
In our experiment, we use a square diamond membrane with a width of $\SI{1}{\milli\meter}$ cut along the $\left<110\right>$ directions and a $(001)$ surface. The diamond membrane has a thickness of $\SI{26}{\micro\meter}$ with the top and bottom layer removed by reactive ion etching to eliminate surface damage caused by polishing \cite{Heupeletch}. SnV centers are generated by implantation of $\SI{65}{\keV}$ tin ions into the membrane, a subsequent annealing to $\SI{1200}{\degreeCelsius}$, and boiling in a tri-acid mixture \cite{Fuchs2021}. The membrane is glued on a silicon substrate using a UV-curing adhesive (NOA63, \textit{Norland products}). Utilizing an adhesive with a significantly different coefficient of thermal expansion compared to that of diamond serves as a simple procedure to intentionally induce strain within the diamond membrane. The silicon substrate has a more than two-fold larger thermal expansion coefficient than diamond.

Our experiments are performed in a home-built dilution refrigerator reaching temperatures as low as $\SI{50}{\milli\kelvin}$. The sample is mounted on a cold finger and accessed through windows in the thermal shields of the cryostat. Optical addressing of the SnV center is achieved through a confocal microscope setup, as described in Appendix \ref{app_sub:setup}.

In Fig. \ref{fig:fig_overwiev} (a), the energy level scheme of a strained SnV center is illustrated. Spin-orbit interaction splits the zero-phonon line (ZPL) into the $E_\mathrm{g,u}^{1/2}$ and $E_\mathrm{g,u}^{3/2}$ doublets, where the $\frac{1}{2}$ and $\frac{3}{2}$ superscripts refer to the total angular momentum, while $\mathrm{g}$ and $\mathrm{u}$ refer to the ground and excited state, respectively \cite{Thiering_magneto_optic}. In photoluminescence (PL) measurements, the effect of strain is observable as an increased splitting between the two doublets in the ground and excited state. The resulting splitting is depicted by blue and red arrows for the C and D transitions, respectively. Zeeman interaction lifts the degeneracy of the doublets and allows for individual addressing of the states by resonant optical excitation or microwave driving. The optically allowed (forbidden) transitions are indicated as black (grey) arrows. The spin qubit is shown as orange-colored energy levels. In Fig. \ref{fig:fig_overwiev} (d) the PL measurements of an unstrained (SnV-D) and a strained (SnV-B) SnV center are shown.
To quantify the strain effect of the adhesive to the SnV centers, we measure the PL spectra of over 400 emitters along the diamond membrane and select the spectra with precisely distinguishable pairs of C and D transitions as depicted in Fig. \ref{fig:fig_overwiev} (e).
With growing distance to the diamond edge, the component $\epsilon_{A_1}$ of the strain tensor, which corresponds to a shift of the ZPL \cite{Meesala_strain}, increases. This is directly observed in the spectra of the SnV centers, where the ZPL at the center of the membrane shifts up by $\SI{1}{\nano\meter}$ compared to the unstrained case. On the other hand, the ground state splitting, dominated by the $\epsilon_{E_{x,y}}$ components \cite{Meesala_strain}, remains nearly constant, as depicted in the lower part of the figure. This is observed for equally prepared samples, although the absolute value of the strain can vary for individual samples, depending, e.g., on the thickness of the adhesive. The combination of the strain components qualitatively explains the behavior of the strain tuned emitters and is consistent with a COMSOL simulation, see Appendix \ref{app:strain}.

\section{\label{sec:sec3}Magneto-Optical properties of the SnV center}

\begin{figure*}
\includegraphics[width=\linewidth]{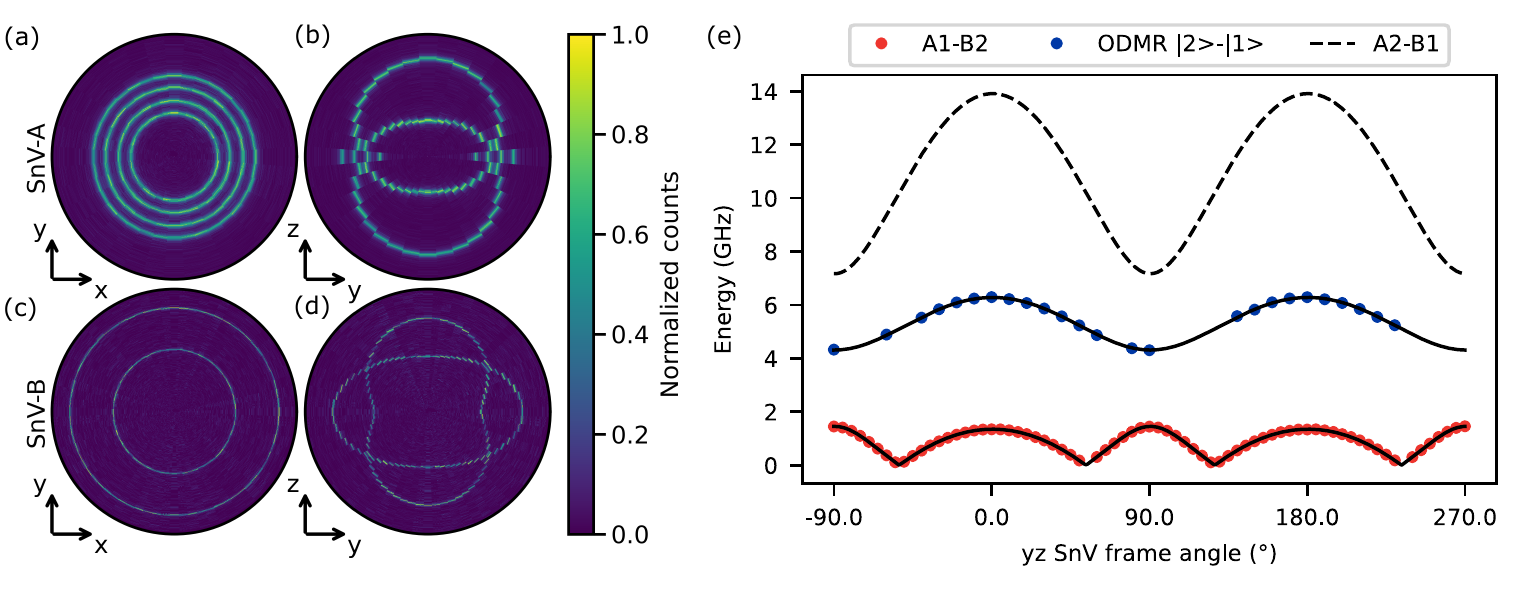}
\caption{\label{fig:fig_ham} Optical properties of the strained SnV center under different magnetic field orientations at \SI{8}{\kelvin}. (a-b) Splitting of the allowed A1 and B2 and forbidden A2 and B1 transitions of SnV-A measured by PLE scans shown in polar coordinates, where the radial axis spans a \SI{2.3}{\giga\hertz} range. Measured with \SI{25}{\mega\hertz\per\second} scanning speed under varying $B$-field orientations with \SI{190}{\milli\tesla} amplitude. (c-d) Splitting of the A1 and B2 transitions of SnV-B measured by PLE scans within a \SI{4.3}{\giga\hertz} range and \SI{50}{\mega\hertz\per\second} scanning speed under varying $B$-field orientations with \SI{190}{\milli\tesla} amplitude. (e) Fit of the electronic energies of SnV-B to the observed optical spin conserving transition A1 and B2 in the SnV-frame (red points). The qubit transition frequencies (blue points), are measured in ODMR scans at \SI{50}{\milli\kelvin}. The spin-forbidden transitions A2 and B1 are not measured due to their large splitting.}
\end{figure*}

\subsection{\label{sec:sec3sub1_Ham_SnV}Hamiltonian of the SnV center}
As a first step, we aim for a comprehensive characterization of the SnV center's optical and magnetic properties. The Hamiltonian for Group-IV defects is originally derived from measurements on silicon-vacancy centers \cite{Hepp_electr_struc_SiV}. Here, we use a Hamiltonian which directly relates to each of the four spin-orbit doublets $E_\mathrm{g,u}^{1/2}$ and $E_\mathrm{g,u}^{3/2}$, resulting in an effective Hamiltonian given by
\begin{multline}
\label{eq:Hamiltonian}
\hat{H}_\mathrm{eff}^\mathrm{g,u} =  -\lambda^\mathrm{g,u} \hat{L}_z \hat{S}_z + \mu_B \left(f_{m_j}^\mathrm{g,u}\right)_{\{\pm\frac{1}{2},\pm\frac{3}{2}\}}\cdot \hat{L}_z B_z \\ + g_s \mu_B \bm{\hat{S}\cdot \hat{B}} + \hat{\Upsilon}_\mathrm{strain}\, .
\end{multline}
The first term accounts for the effective spin-orbit splitting. The second term corresponds to the orbital Zeeman effect that contains the Bohr magneton $\mu_B$ and the quenching factor $f_{m_j}^\mathrm{g,u}$ that corrects the orbital Landé g-factor, i.e. describes the smaller splitting opposed to the expected splitting for a quantized magnetic moment. This quenching factor can be separated into two parts $f_{m_j}^\mathrm{g,u} = p_{m_j}^\mathrm{g,u} \cdot g_l^\mathrm{g,u}$, where $p_{m_j}^\mathrm{g,u}$ is the contribution by the dynamic Jahn-Teller effect (DJT) known as the Ham effect, and $g_l^\mathrm{g,u}$ is the Steven's orbital quenching factor, arising from lowered symmetry of the system \cite{Thiering_magneto_optic}. Each magnetic quantum number $m_j = \pm\frac{1}{2}$ and $m_j = \pm\frac{3}{2}$ corresponds to one of the electronic states of the spin-orbit doublets split by the applied external magnetic field. Attributing a separate quenching factor to each single orbital with corresponding $m_j$, instead of using an averaged quenching with an asymmetry \cite{Rosenthal, Guo, Thiering_magneto_optic}, allows for a precise determination of the orbital quenching in presence of different strain environments.
For instance, for an unstrained emitter, the qubit frequency, that corresponds to the splitting of the $E_\mathrm{g}^{3/2}$ doublet, is purely determined by the $m_j=\pm\frac{3}{2}$ orbitals. Hence, an average quenching factor $f^\mathrm{g}$ for the two ground state doublets cannot be extracted, as it lacks a meaningful physical basis. However, the $m_j = \frac{1}{2}$ substates contribute for strained emitters, as we show in the derivation of the Hamiltonian in Appendix \ref{app:Hamiltonian}, and can be extracted as a consequence.
We use four separate sets of data, each from different emitters, with unstrained (SnV-D), low strain (SnV-A), and high strain (SnV-B \& SnV-C) properties and follow a step by step approach in our fitting routine to determine the quenching factors for all ground and excited state doublets. The third term in the effective Hamiltonian in Equation \ref{eq:Hamiltonian} is the electron Zeeman contribution with the electron $g$-factor ($g_s=\num{2}$) and the Pauli-matrix vector $\bm{\hat{S}}=\frac{1}{2}(\hat{\sigma}_x,\ \hat{\sigma}_y,\ \hat{\sigma}_z)^\mathrm{T}$. Finally, the fourth gives the strain contribution $\hat{\Upsilon}_\mathrm{strain}$, which contains
the diagonalized strain contributions $\alpha$, for ground and
excited state, respectively. For an in-detail discussion and matrix form representations we refer to Appendix \ref{app:Hamiltonian}.

\subsection{\label{sec:sec3sub1_3D_mag}Analysis of magnetic-field dependent transitions}

In the following, we discuss the influence of strain on the electronic spin's optical addressability and fit the Hamiltonian to our measurements. In Fig. \ref{fig:fig_ham} (a)-(b) ((c)-(d)) the dependence of the optical transitions on the magnetic field orientation of the low (high) strain SnV-A (SnV-B) are presented. The image shows photoluminescence excitation (PLE) measurements of the spin-conserving optical transitions A1 and B2 at a temperature of \SI{8}{\kelvin}. The magnetic field orientation is given in polar coordinates within the respective frame of the SnV center. The radial axis corresponds to the frequency of the PLE scans at fixed polar coordinates. This data is obtained by preceding measurements of the A1 and B2 transitions in the lab frame under an external magnetic field rotation. With the SnV quantization axis obtained in this way, we acquire {\it xy} and {\it yz} rotation maps, i.e. the external magnetic field is rotated in both polar $\theta$ and azimuthal $\phi$ directions in the frame of the investigated emitter. During the measurement, the magnetic field magnitude is held constant at $B_\mathrm{dc} = \SI{190}{\milli\tesla}$. The radial axis in Fig. \ref{fig:fig_ham} spans over \SI{2.3}{\giga\hertz} and \SI{4.3}{\giga\hertz} of the excitation laser frequency for SnV-A and SnV-B, respectively.
We note that the absolute magnitude of the external magnetic field plays a crucial role for the precision of the determined quenching factors. Thus, we include a thorough uncertainty analysis of the calibration process of the magnetic field strength in Appendix \ref{app:Hamiltonian}.
Scanning in the $xy$-plane as seen Fig. \ref{fig:fig_ham} (a) and (c), i.e. perpendicular to the quantization axis, serves to validate the experimental precision and to gain insight into the strain magnitude of the orbital ground and excited states. In the presence of strain, the degeneracy of these transitions is lifted. This can be directly observed by the separation of the optical lines.
Furthermore, for SnV-A in Fig.~\ref{fig:fig_ham} (a), along with the two allowed transitions A1 and B2, the two forbidden transitions A2 and B1 can be seen, as a result of spin-mixing under these conditions.
The energy separation of the A1 and B2 transitions measured for magnetic field rotations in the $yz$-plane allows further constraint on the Hamiltonian. Fig. \ref{fig:fig_ham} (e) exemplarily shows this difference for SnV-B as a function of the polar angle in the {\it yz}-plane using red points. Additionally, the qubit transitions are determined from optically detected magnetic resonance (ODMR) measurements and depicted as blue points. The data can be fully described by the Hamiltonian given in Equation \ref{eq:Hamiltonian}.

In our procedure, we first determine the ground state splitting of an unstrained emitter (SnV-D) through PL measurements, as illustrated in Fig. \ref{fig:fig_overwiev} (d). The absence of strain is validated by the preserved degeneracy of the A1 and B2 transitions under a perpendicular magnetic field, see Appendix \ref{app:derivation_ham} and Fig. A\ref{appfig:Polar_SnV_d}.
We determine the spin-orbit interaction for the ground state to be $\lambda^\mathrm{g} = \SI{822\pm 1}{\giga\hertz}$, matching previous reports for high-temperature treated samples \cite{Narita_idtcal_photon}.
For strained SnV centers, like the one depicted in Fig. \ref{fig:fig_ham}, the ground state strain $\Upsilon^\mathrm{g}$ is related to the ground state splitting by the relation $\Delta_\mathrm{g} = \sqrt{{(\lambda^\mathrm{g})}^2 + 4{(\Upsilon^\mathrm{g})}^2}$.

We use the determined spin-orbit coupling $\lambda^\mathrm{g}$ as input to find the ground state strain magnitude by matching it to the minimum of the qubit transitions $\nu_\mathrm{qubit}(\theta)$ at $\theta = \SI{90}{\degree}$, as this frequency is solely depending on $\Upsilon^\mathrm{g}$ for a given magnetic field strength. For the excited state strain $\Upsilon^\mathrm{u}$, we fit the Hamiltonian at the same angle $\theta = \SI{90}{\degree}$ to the measured allowed transition splitting, assuming a spin-orbit coupling for the excited state of $\lambda^\mathrm{u}=\SI{3000}{\giga\hertz}$ \cite{Görlitz_2020, Thiering_magneto_optic}.

In the next step, we determine $f_{\frac{3}{2}}^\mathrm{g}$ by fitting the qubit transitions of a low strain emitter (SnV-A) with $\Upsilon^\mathrm{g} = \SI{35\pm 1.5}{\giga\hertz}$. This value is completely independent from all other quenching factors, when fitted to the qubit transitions.
By fixing this value, we can now determine $f_{\frac{3}{2}}^\mathrm{u}$, by fitting to the allowed transition splitting of SnV-A.
The $f_{\frac{1}{2}}^\mathrm{g,u}$ values cannot be determined in the case of low strain, as their influence is negligible due to the absence of orbital mixing. However, in the case of moderate and high strain where $\lambda \ll \alpha$ does not apply, the influence of the quenching factors with $m_j = \pm \frac{1}{2}$ increases.
Hence, we use the obtained values for $f_{\frac{3}{2}}^\mathrm{g,u}$ from SnV-A to further determine the $f_{\frac{1}{2}}^\mathrm{g}$ value by fitting to the qubit transitions of the highly strained SnV-B.
Here, the $f_{\frac{1}{2}}^\mathrm{g}$ sublevel of the ground state significantly contributes to the energy levels of the qubit in the diagonalized Hamiltonian.  It is, however, insensitive to $f_{\frac{1}{2}}^\mathrm{u}$ from the excited state. Lastly, we determine the remaining value by fitting to the allowed transitions of SnV-B.
To validate our fitting procedure, we use the dataset obtained by an additional emitter (SnV-C) and compare the measurement data with the fitting parameters, see Fig. A\ref{appfig:Ham_fit_all} (e-f). The final fitting parameters and the corresponding errors are shown in Table \ref{tab:Ham_fit_f_pars}.

\begin{table}[b]
\caption{\label{tab:Ham_fit_f_pars}
Parameters of the SnV center Hamiltonian. The spin-orbit splitting $\lambda^\mathrm{g}$ is measured by PL measurements from an unstrained SnV center (SnV-D). For the magnetic field magnitude values, we estimate an error of 0.5\,\%. The strain components and orbital quenching factors are obtained by a fit to the electron spin Hamiltonian.}
\begin{ruledtabular}
\begin{tabular}{cccc}
Parameter & Value & Fitted? & Source \\
\hline
$\gamma_l$ & \SI{14}{\giga\hertz\per\tesla} & - & \cite{gammal}\\
$\gamma_s$ & \SI{28}{\giga\hertz\per\tesla} & - & \cite{gammae}\\
$\lambda^\mathrm{g}$ & \SI{822 \pm 1}{\giga\hertz} & yes & SnV D\\
$\lambda^\mathrm{u}$ & \SI{3000}{\giga\hertz} & - & \cite{Görlitz_2020, Thiering_magneto_optic}\\
$f_{\frac{3}{2}}^\mathrm{g}$ & \num{0.268 \pm 0.013} & yes & SnV A\\
$f_{\frac{1}{2}}^\mathrm{g}$ & \num{0.251 \pm 0.012} & yes & SnV A\\
$f_{\frac{3}{2}}^\mathrm{u}$ & \num{0.486 \pm 0.015} & yes & SnV B\\
$f_{\frac{1}{2}}^\mathrm{u}$ & \num{0.500 \pm 0.008} & yes & SnV B\\
\hline
 & SnV A  & SnV B & SnV C \\
\hline
$\Upsilon^\mathrm{g}$  & \SI{35.0 \pm 1.5}{\giga\hertz} & \SI{577.3\pm 3.4}{\giga\hertz} & \SI{530.0 \pm 4.0}{\giga\hertz}\\
$\Upsilon^\mathrm{u}$  & \SI{60 \pm 7.0}{\giga\hertz} & \SI{961.9 \pm 0.49}{\giga\hertz} & \SI{921.4 \pm 3.2}{\giga\hertz}\\
$B_{\parallel}$  & \SI{193.44 \pm 0.97}{\milli\tesla} & \SI{193.46 \pm 0.97}{\milli\tesla} & \SI{193.47 \pm 0.97}{\milli\tesla}\\
$B_{\perp}$  & \SI{193.48\pm 0.97}{\milli\tesla} & \SI{189.03 \pm 0.95}{\milli\tesla} & \SI{193.45 \pm 0.97}{\milli\tesla}\\
$\delta\theta$  & \SI{-0.08\pm 0.02}{\degree} & \SI{-0.54 \pm 0.02}{\degree} & \SI{-0.46 \pm 0.18}{\degree}\\
\end{tabular}
\end{ruledtabular}
\end{table}

\section{\label{sec:sec4}Coherent Control}
\subsection{\label{sec:sec4sub1_Nb_CPW}Superconducting coplanar waveguide}

\begin{figure}
\includegraphics[width=\linewidth]{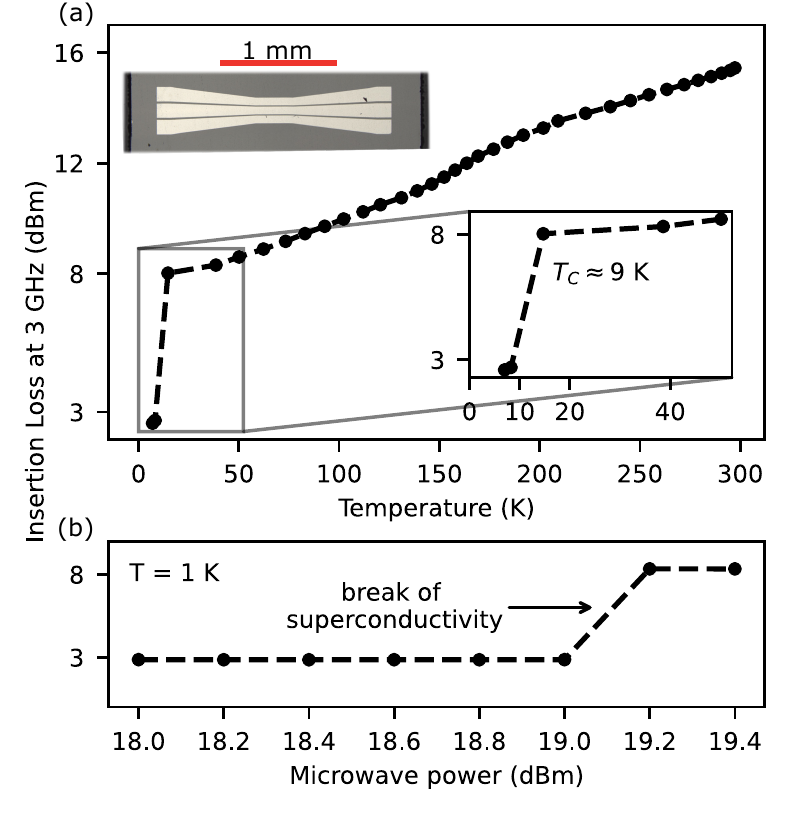}
\caption{\label{fig:fig_waveguide} Properties of superconducting niobium waveguides. (a) Insertion loss (S21) of a superconducting CPW at a frequency of $\SI{3}{\giga\hertz}$. The enhanced transmission at $\lesssim \SI{9}{\kelvin}$ is caused by the superconducting phase transition and the vanishing resistance. The CPW for the reference measurements is depicted in the inset in (a) and was fabricated on a bulk diamond. (b) Break of superconductivity at a power of $\SI{19}{\dBm}$.}
\end{figure}

\begin{figure*}
\includegraphics[width=\linewidth]{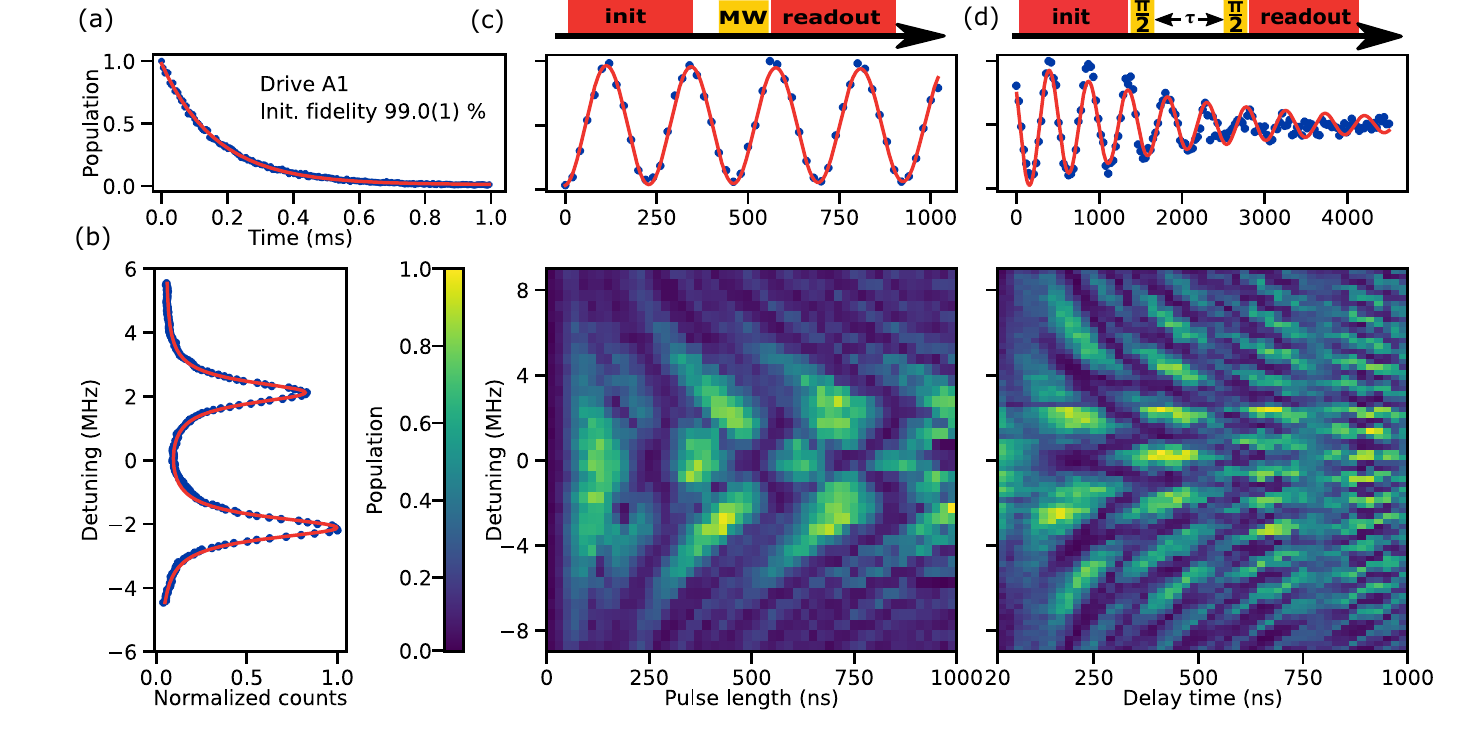}
\caption{\label{fig:fig_coh_control} Coherent manipulation of the electron spin at \SI{50}{\milli\kelvin}. (a) Selective spin initialization of A1 with an initialization fidelity of $\SI{99.0 \pm 0.1}{\%}$. (b) ODMR with continuous wave excitation. Two resonances are observed due to coupling to a proximal nuclear spin. The red line corresponds to a double Lorentzian fit with center frequencies $\SI{3144.34 \pm 0.22}{\mega\hertz}$ and $\SI{3148.58 \pm 0.26}{\mega\hertz}$ and a linewidth of $\SI{838.65\pm 0.82}{\kilo\hertz}$. (c) Exemplary Rabi oscillation. Top: Scheme for resonant initialisation, coherent driving and resonant readout. Single oscillation with a Rabi frequency of $\Omega/2\pi=\SI{4.31 \pm 0.28}{\mega\hertz}$ at $\SI{14}{\dBm}$. Bottom: 2D Chevron pattern with varying detuning showing a beating due to the hyperfine structure. (d) Ramsey measurement at the center frequency. Data points can be fitted by a sinusoidal with an exponential decay of $\SI{2.1 \pm 0.2}{\micro\second}$. Bottom: 2D Ramsey pattern.}
\end{figure*}

With the collected understanding of the spin Hamiltonian, we redirect our attention to the coherent manipulation of the electron spin. In recent studies on microwave spin control of SnV center qubits in diamond \cite{Guo, Rosenthal, Debroux2021}, drive-induced heating has been identified as a significant limitation. This plays a pivotal role in limiting the spin lifetime $T_1$ and increasing the infidelity of quantum operations.

We briefly examine these limitations arising from intrinsic Ohmic losses by fabricating a normal conducting microwave coplanar waveguide made of gold on a diamond sample. We study an SnV center $\approx \SI{10}{\micro\meter}$ away from the central conductor and measure the photon counts under resonant excitation. After initialization in the off-resonant spin state the emitter remains dark at low temperature, given the long spin relaxation time.
We then apply an off-resonant microwave pulse of \SI{10}{\milli\second} length and observe that already at moderate microwave powers around $\SI{0}{\dBm}$, the photon count instantly increases and follows a non-trivial behavior for different microwave powers. We interpret the increasing countrate as a reduced $T_1$ time originating from an instant heating effect upon the application of microwaves, followed by diffusion of the locally induced heat into the cryostat.
The precise local temperature is impossible to determine with conventional thermometers, as the heat dissipates over the large distance from the SnV center towards the thermometer. However, heat is instantly felt by the SnV center and detectable by observing the count rate under resonant excitation of one spin state. From our measurements we can deduce that the temperature dependent lifetime $T_1$ is reduced by several orders of magnitude and the count rate increases by thermal re-population of the continuously pumped readout state. While this effect can be interesting for temperature sensing at temperatures below $\SI{4}{\kelvin}$, it poses a bottleneck for microwave control. For more details and the measurement data we refer to Appendix \ref{app:heating_effects}.

In order to overcome this obstacle arising from microwave-line heating, we make use of the absence of Ohmic losses of superconducting metals \cite{SCRes} and fabricate a coplanar waveguide (CPW) made from niobium. Niobium, with a high critical temperature of $\sim \SI{9.2}{\kelvin}$ \cite{Niob_Tc}, is a suitable material for microwave control even at elevated temperatures that are commonly achieved in bath or cryogen-free cryostats.
To evaluate the CPW for our microwave control measurements, we first determine the properties of superconducting waveguides (SC-CPW). On a bulk diamond substrate, a continuous CPW is fabricated via all-optical lithography, as is shown in the inset in Fig. \ref{fig:fig_waveguide} (a). The CPW has a thickness of $\SI{50}{\nano\meter}$ and is fabricated by electron-beam evaporation.
In Fig. \ref{fig:fig_waveguide} (a), we depict the measured insertion loss (S21) of the coaxial lines and the waveguide at a frequency of $\SI{3}{\giga\hertz}$ as a function of the temperature, that is acquired during the cooldown of the cryostat. With decreasing temperature, the total resistance decreases continuously, apart from a sharp drop at around $\SI{9}{\kelvin}$, when the niobium film is entering the superconducting state. The residual losses arise from impedance mismatch and the normal-conducting coaxial lines inside the cryostat.
The electric current through the waveguide is proportional to the effective $B_\mathrm{ac}$ and proportional to the square root of the applied microwave power. For that reason, we evaluate the superconductor in terms of the maximum applicable power. Fig. \ref{fig:fig_waveguide} (b) shows the transition from the superconducting to the normalconducting state at a power of $\SI{19}{dBm}$.

For the microwave spin control measurements, we use a simple SC-CPW design with a short circuit at the waveguide's end, see Fig. \ref{fig:fig_overwiev} (c). We simulate the waveguide in the presence of the dielectric environment and analytically determine the required size of the CPW gaps. This analysis ensures that the characteristic impedance matches the impedance of $Z_0 = \SI{50}{\ohm}$ of the transmission lines into the cryostat, avoiding reflections through impedance mismatch. By taking the dielectric properties of the silicon substrate, the polymer, and the diamond membrane, into account \cite{CPW_wiley}, we determine the size of the gaps between the conductors to $\SI{24}{\micro\meter}$ for conductors with a width of $\SI{150}{\micro\meter}$.

\begin{figure}
\includegraphics[width=\linewidth]{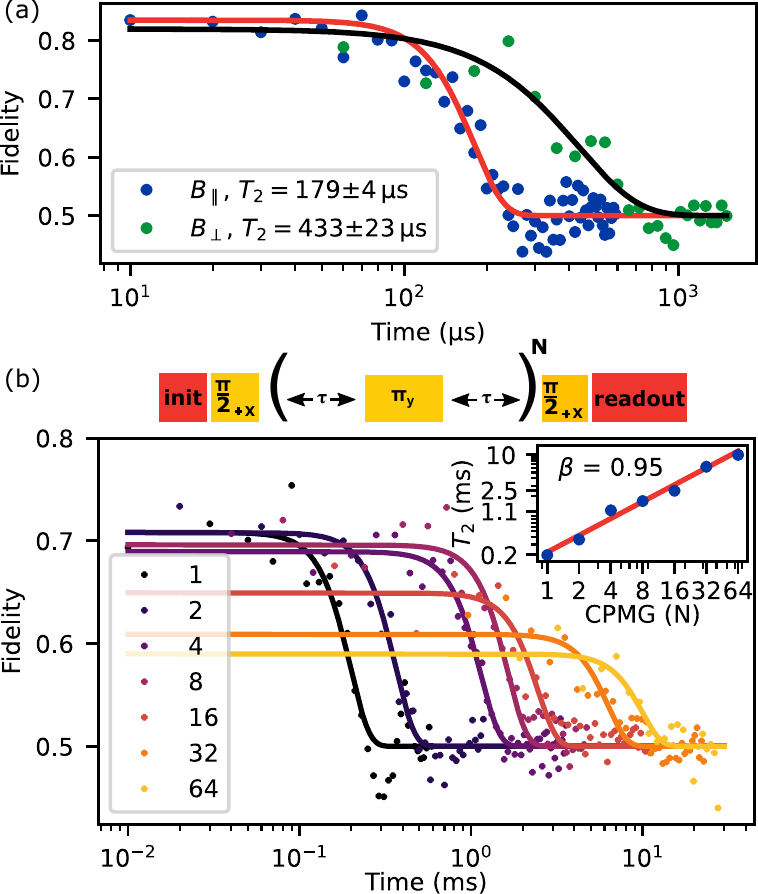}
\caption{\label{fig:fig_cpmg} Coherence measurements. (a) Hahn echo measurements under parallel and perpendicular applied magnetic field. (b) CPMG sequence with varying number $N$ refocusing pulses. The data is fit to stretched exponential envelopes with $e^{-(t/T_2)^{\xi}}$ with $\xi=4$. Inset shows a double-logarithmic plot of  a power-law behavior $T_2 \propto N^{\beta}$  with a $\beta = 0.95$.}
\end{figure}

\subsection{\label{sec:sec4sub1_MW_spin_control}Microwave spin control}
In a next step, we make use of the SC-CPW for coherent spin control. By applying an external magnetic field, the lower $E_\mathrm{g}^{3/2}$ doublet splits into the spin states $\ket{1}$ and $\ket{2}$ that can be used as a qubit. Here, we demonstrate coherent control of the SnV under a parallel magnetic field of $B_\parallel = \SI{96}{mT}$ at a temperature of $\SI{50}{\milli\kelvin}$.
The initialization of the electron spin is achieved by resonantly driving one of the allowed optical transitions, either A1 or B2 as seen in Fig. \ref{fig:fig_overwiev} (a).
We achieve an initialization fidelity of up to $F=\SI{99.0 \pm 0.1}{\%}$, which is limited by the dark counts of the single photon detector, as shown Fig. \ref{fig:fig_coh_control} (a). The optical cyclicity is particularly high for parallel field orientation, yielding a high readout signal but also increasing the initialization time. The latter can be adjusted by modifying the excitation power of the resonant laser beam.
We observe a highly stable negative charge state of the SnV center and thus, do not implement an off-resonant repump pulse in our control sequences. If occasionally necessary, the charge is regained by a $\SI{532}{\nano\meter}$ repump pulse with excitation power below $\SI{100}{\nano\watt}$ to avoid spectral diffusion due to changes in the charge environment \cite{Görlitz_2020}.
We determine our qubit frequency via ODMR measurements as shown in Fig. \ref{fig:fig_coh_control} (b) by using continuous microwave chirps. The microwave frequency is swept within a fixed time of \num{10} to $\SI{20}{\milli\second}$ under continuous resonant optical driving using power in the order of \SI{1}{\nano\watt} and repeated until the desired signal-to-noise ratio is gained.
To achieve high spectral resolution, we use low microwave amplitudes at insertion powers as low as $\SI{-40}{dBm}$ (\SI{100}{\nano\watt}) into the cryostat. We observe two resonances for parallel magnetic field orientation at $\SI{3144.34 \pm 0.22}{\mega\hertz}$ and $\SI{3148.58 \pm 0.26}{\mega\hertz}$ with a linewidth of $\SI{838.65\pm 0.82}{\kilo\hertz}$. The separation indicates strong coupling to a proximal spin.
From the beating in the Spin-Echo measurements, see Appendix \ref{appfig:Echo_all}, we infer that the splitting originates from a single $^{13}\mathrm{C}$ nuclear spin. Such coupled nuclear spins can serve as a long-lived quantum memory \cite{Metsch2019, Parker2024}.

Fig. \ref{fig:fig_coh_control} (c) illustrates a Rabi measurement at $\SI{14}{\dBm}$ insertion power, showing a Rabi frequency of $\Omega/2\pi = \SI{4.31 \pm 0.28}{\mega\hertz}$ that corresponds to a $\pi$ pulse of $\SI{115}{\nano\second}$. The full contrast and absence of decay illustrates the high fidelity ($F=\SI[parse-numbers=false]{99.1(+0.3)(-2.3)}{\percent}$) of the spin manipulation. In the Chevron pattern of the Rabi oscillation, the coupling to the nuclear spin is directly visible by the presence of two resonances where the Rabi frequency is minimal. However, the measured pattern is not reproduced by the sum of two independent detuned Rabi oscillations as one may expect for a thermal nuclear spin. The deviation can be explained by electron-nuclear correlations that are introduced by driving the electron spin under conditions where the hyperfine coupling to the nuclear spin is of the same order of magnitude as the qubit's Rabi frequency \cite{JelezkoHHDR}. This mechanism can be used for coherent nuclear spin control \cite{Metsch2019}.

Performing free-induction decay measurements allows to probe the dephasing time of the system. From the envelope of a Ramsey measurement tuned to the center between the two resonances, as pictured in Fig. \ref{fig:fig_coh_control} (d), we determine a dephasing time $T_2^{*} = \SI{2.1 \pm 0.2}{\micro\second}$ of the qubit, which is in the same order as for isotopically pure overgrown diamond \cite{Guo}. As we detune the driving frequency relative to the two transitions, the signal oscillates and forms a crown like pattern, with instances of a full loss of coherence again stemming from the coupling to the proximal nuclear spin.

\subsection{\label{sec:sec4sub2_DD}Dynamical Decoupling}
The coherence time $T_2$ can be increased significantly compared to $T_2^{*}$ by making use of echo techniques. First, we measure the coherence time for a Hahn echo measurement. We drive resonantly to one nuclear spin resonance with a microwave power of $\SI{10}{\dBm}$, resulting in a Rabi frequency of $\SI{2.5}{\mega\hertz}$. We determine a coherence time $T_2 = \SI{179 \pm 4}{\micro\second}$ ($T_2 = \SI{433 \pm 23}{\micro\second}$) for parallel (perpendicular) field orientation, see Fig. \ref{fig:fig_cpmg} (a).
The higher coherence time for a perpendicular field orientation can be attributed to the decoupling from the electron spin bath \cite{Rosenthal}. To further extend the coherence time, we incorporate the Carr-Purcell-Meiboom-Gill (CPMG) sequence that consists of multiple refocusing $\pi$ pulses, shifted in phase with respect to the first $\frac{\pi}{2}$ pulse.
As a perpendicular magnetic field orientation results in a very small cyclicity of the readout state and thus requires many repetitions to achieve reasonable SNRs, we measure all CPMG sequences with a parallel field orientation.
Fig. \ref{fig:fig_cpmg} shows the measurements for up to 64 refocusing pulses, increasing the coherence time to $T_{2,\mathrm{CPMG}64} = \SI{10 \pm 1}{\milli\second}$, which is about six-fold improved compared to earlier work \cite{Guo}. The fidelity of the decoupling decreases for higher number of pulses, as more and more pulse errors due to the coupling of the proximal nuclear spin are added.
However, we determine the scaling of $T_2$ with the pulse number and find an almost linear scaling $T_2\propto N^\beta$ with a $\beta= \SI{0.95 \pm 0.09}{}$. We attribute the large scaling factor to strongly reduced Ohmic losses by using a superconducting waveguide.

\section{\label{sec:sec5}Conclusion}
In this work, we demonstrate extended control over the optical and magnetic levels of the SnV center electron spin, thereby contributing to a deeper understanding of the these properties. We show a simple procedure to incorporate strain into the diamond to allow for microwave control. We measure 2D rotation maps of the optical transitions under varying magnetic field orientations to determine the SnV axis with $<\SI{1}{\degree}$ uncertainty.
We develop a fitting procedure for the optical and microwave transitions to the Hamiltonian of the electron spin, using the explicit doublets for each total angular momentum, which allows for a qualitative determination of the orbital quenching factors. We use niobium to avoid drive induced heating and obtain long coherence times of up to $T_{2} = \SI{10 \pm 1}{\milli\second}$ in a CPMG sequence. The almost linear scaling of the coherence time with the number of pulses shows potential for further improvement.
For higher fidelity of the signal, pulse errors can be minimized by using optimized control pulses \cite{optimalControlJulich, HalfmannUDD}, shorter $\pi$-pulses, and initialization of the proximal nuclear spin \cite{JelezkoHHDR}.
Rabi frequencies can be further enhanced by optimizing the design of the coplanar waveguide and increasing the critical current density. We have shown, that the use of superconducting waveguides allows for an efficient control of the SnV center electron spin, achieving long coherence times.
Our careful analysis of the electron spin Hamiltonian provides valuable insights for further investigations on the response of the SnV center under coherent driving \cite{Pieplow_strain}. The observed coupling to a proximal $^{13}\mathrm{C}$ nuclear spin, promises future possibilities to manipulate and store quantum information into its long-lived spin-degree.

\begin{acknowledgments}
We thank Tim Schröder, Gregor Pieplow, Mohamed Belhassen and Dennis Rieger for helpful discussions, Eckhard Wörner from Diamond Materials for cutting and notching the sample, and Jonathan Körber for support in the early stage of the experiment.

This work was partly supported by the the German Federal Ministry of Education and Research (Bundesministerium für Bildung und Forschung, BMBF) within the project QR.X (Contracts No. 16KISQ004 and No. 16KISQ005), SPINNING (Contract No. 13N16211), the Deutsche Forschungsgemeinschaft (DFG) through CRC TRR 288 - 422213477 "ElastoQMat" (Project A08), the Kompetenzzentrum Quantencomputing Baden-Württenberg (Contract QC-4-BW), and the Karlsruhe School of Optics and Photonics (KSOP).
\end{acknowledgments}

\nocite{*}

\bibliography{apssamp}

\clearpage
\appendix
\onecolumngrid 
\section{\label{app:Exp_det}Experimental details}
\subsection{\label{app_sub:sample_fab}Sample fabrication}
The sample used in these experiments is a electronic grade diamond membrane from \textit{Element Six} with dimensions $2\times2\times\SI{0.04}{\milli\meter}$ and an initial surface roughness of \SI{4}{\nano\meter} (\SI{2.3}{\nano\meter}) on side A (side B). Both sides are strain relief etched by reactive ion etching. The parameters for the etching procedure are as follows
\begin{itemize}
    \item[--] side A:\\
    \SI{50}{\min} Ar/Cl${}_2$ + 5 x (\SI{7}{\min} Ar/Cl${}_2$ + \SI{13}{\min} O${}_2$),
    \item[--] side B:\\
    \SI{20}{\min} min Ar/Cl${}_2$ + 5 x (\SI{7}{\min} min Ar/Cl${}_2$ + \SI{15}{\min} O${}_2$) + \SI{10}{\min} Ar/Cl${}_2$ + \SI{15}{\min} O${}_2$,
\end{itemize}
to remove damage by the polishing. For further information see similar works, e.g. \cite{Heupeletch}. In total, a thickness of \SI{7}{\micro\meter} was removed on each side, leading to a surface roughness of \SI{3}{\nano\meter} (side A) and \SI{2}{\nano\meter} (side B).

As side B has a lower surface roughness, this side is used for implantation of tin ions. The isotope ${}^{116}\mathrm{Sn}$ is implanted with a fluence of \SI{1e9}{\per\centi\meter\squared} and an energy of \SI{65}{\kilo\electronvolt}. The sample is annealed post implantation for \SI{4}{\hour} at \SI{1200}{\degreeCelsius} at a pressure of $<\SI{1e-6}{\milli\bar}$. To clean the surface, the sample is boiled in a tri-acid mixture (1:1:1, nitric acid:sulfuric acid:perchloric acid) at \SI{400}{\degreeCelsius}. Afterwards, the sample is laser-cut into \num{4} pieces by \textit{Diamond Materials}. At last, the membrane is baked on a hotplate for \SI{6}{\hour} at \SI{450}{\degreeCelsius} and cleaned in boiling piranha acid. 

For the CPW fabrication, we glue one piece of the diamond on a $4\times\SI{4}{\milli\meter}$ Silicon wafer with a UV-curing adhesive. We use the AZ5214E photoresist in our all-optical lithography process and deposit a $\SI{50}{\nano\meter}$ thick niobium layer by electron-beam evaporation. For the lift-off procedure we use acetone and undertake a final cleaning step with water.

\subsection{\label{app_sub:setup}Setup}
The data in this work is measured in a home-built dilution cryostat with a base temperature of \SI{50}{\milli\kelvin}. The sample is mounted on a copper coldfinger that is thermalized to the to the \si{\milli\kelvin} plate with silver strands. The coldfinger is mounted on piezoelectric steppers (2 ANPx101/RES/LT, ANPz101/RES/LT, ANC300-controller, \textit{Attocube}). Optical excitation and readout is performed in a home-built confocal setup, see Fig. \ref{app_sub:setup}.

Resonant laser pulses are generated by a tunable continuous-wave laser source (C-WAVE vis, \textit{Hübner}), locked to a wavemeter (WS7, \textit{HighFinesse}) in addition with an acousto-optical modulator (3200-1214, \textit{G\&H}) . The trigger signals are generated by a fast TTL logic (ADwin-Pro II, \textit{Jäger}). The microwave signal is generated by an arbitrary waveform generator (AWG70001A, \textit{Tektronix}). Charge repump of the SnV center is applied by a continuous-wave \SI{532}{\nano\meter} laser (Ventus MPC6000, \textit{Laser Quantum}) and triggered by a TTL logic (Adwin Gold II, \textit{Jäger}). The magnetic field of up to \SI{1.5}{\tesla} is created by a home-built superconducting 3D vector magnet around the microscope objective (MPLN100X, \textit{Olympus}). The excitation laser is scanned using a galvo mirror system (GVS212, GPS011-power supply \textit{Thorlabs}). Fluorescence is separated by a 90:10 beamsplitter (BSN10, \textit{Thorlabs}) and additional filters (BLP01-532R-25, FF01-593/LP-25, LP01-633R-25, \textit{Semrock}). Finally, the signal is focused through a pin-hole onto an avalanche photo-detector (SPCM-AQRH-16, \textit{Excelitas}). PL spectra are taken with a fiber-coupled spectrometer (SP-2500i, \textit{Princeton Instruments}). 

\begin{figure}
\includegraphics[width=0.9\textwidth]{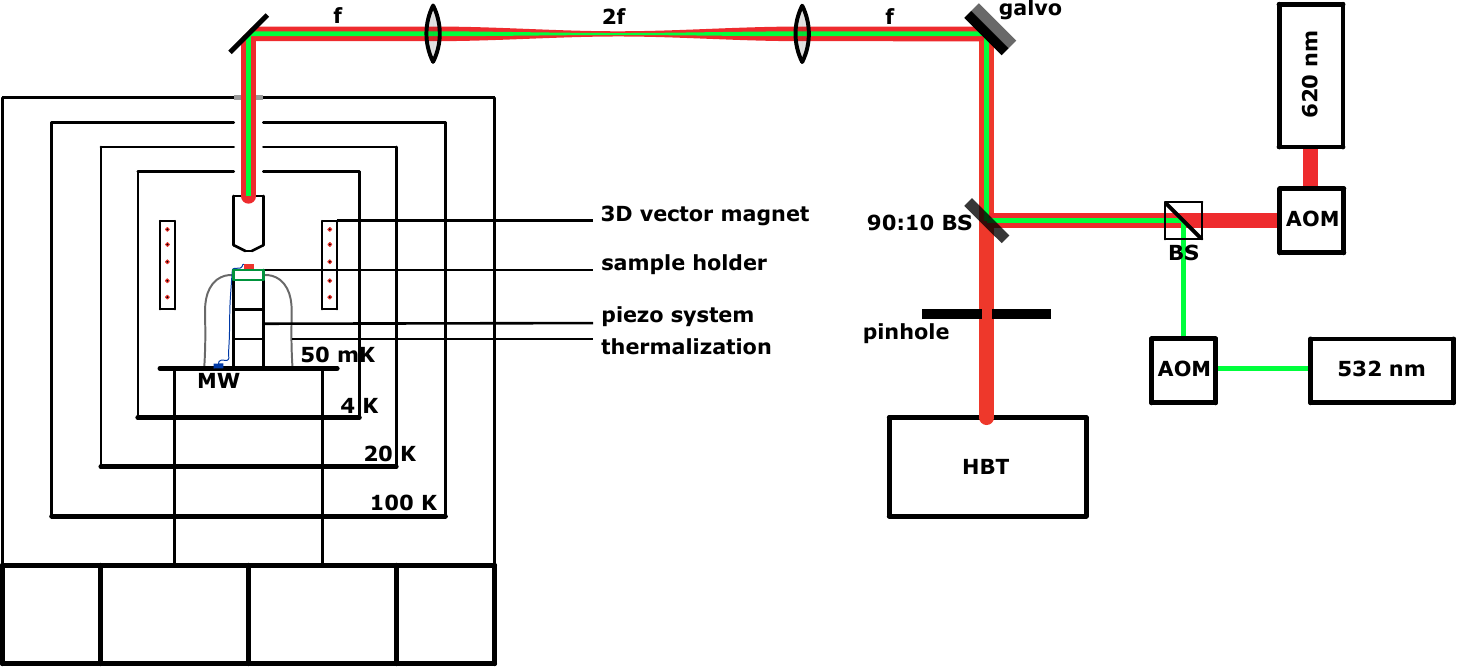}
\caption{\label{appfig:setup} Schematic of the experimental setup. Optical addressing and readout of the SnV centers is performed by a home-built confocal setup, depicted by the red (green) beam paths for resonant (off-resonant) excitation. Side-band fluorescence is separated by a 90:10 beamsplitter and long-pass filters (not depicted) and focused through a pinhole onto avalanche photo-detectors arranged in a Hanbury, Brown-Twiss-setup. The sample is mounted on a piezo stack inside a home-built dilution cryostat with base-temperature of \SI{50}{\milli\kelvin}. Scanning over the sample is achieved by two galvo mirrors arranged in a $4f$ setup, where the objective is mounted onto the \SI{4}{\kelvin} stage. Magnetic fields are applied by a superconducting 3D vector magnet around the focal point of the objective.}
\end{figure}

\subsection{\label{app_sub:MW_caharacterization}Microwave power characterization}
We use coaxial transmission lines for microwave delivery within the cryostat, extending up to the mK-stage. To the sample, we use a custom made CPW fabricated on a flexible printed-circuit board, featuring $\SI{18}{\micro\meter}$ thick copper conductors and a $\SI{100}{\micro\meter}$ thick polyimide dielectric.

\begin{figure}
\includegraphics[width=0.9\textwidth]{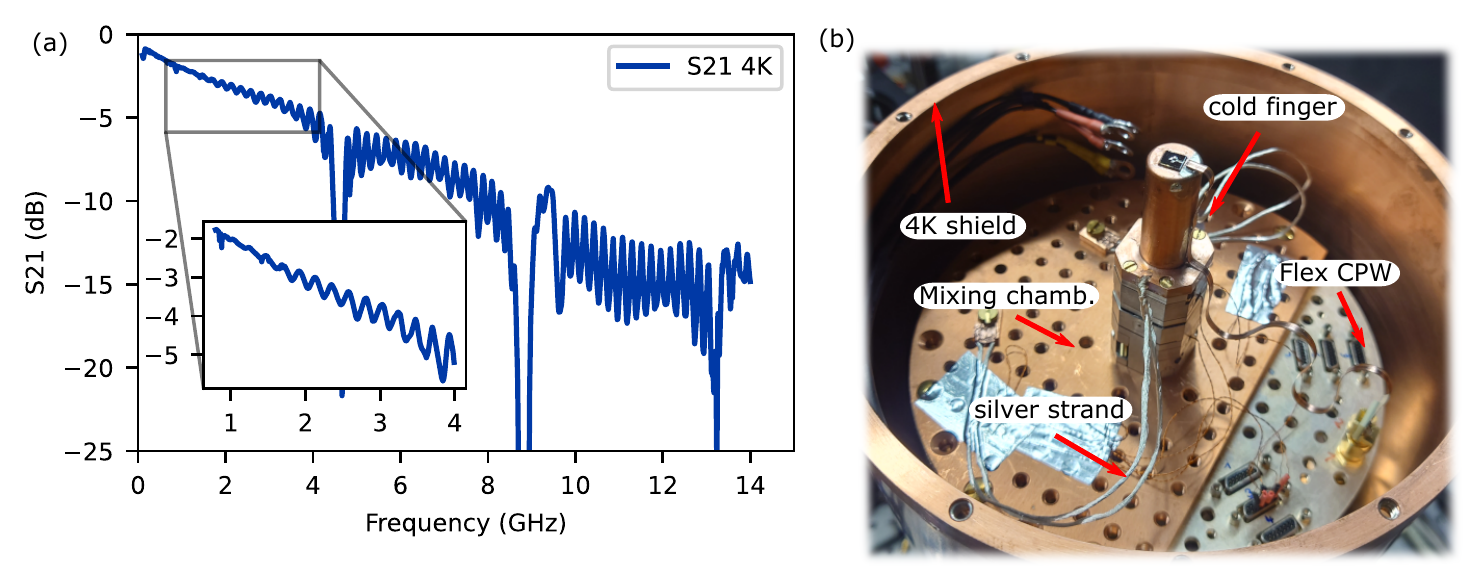}
\caption{\label{appfig:S21} Insertion loss through the cryostat and mixing chamber. (a) S21 at $\SI{4}{\kelvin}$ from the coaxial input of the cryostat to the output including the flexible CPW cable. (b) Mixing chamber of the cryostat. Microwaves reach the sample on the coldfinger via bonding wires from the flexible CPW to the SC-CPW on the sample.}
\end{figure}

The insertion loss (S21) through the cryostat, including the flexible CPW cable, is shown in Fig. A\ref{appfig:S21} (a). The total loss at $\SI{3}{\giga\hertz}$ amounts to approximately $\SI{4}{\dB}$. Due to the symmetry of the setup we expect half the total losses at the sample position. The dips in transmission at $\approx 4.3$ and $\SI{8.6}{\giga\hertz}$ are caused by radiation losses arising from standing waves between the evenly spaced vias with $\SI{1}{\milli\meter}$ distance ($\frac{\lambda}{4}-\mathrm{radiation}$). The vias connect the CPW ground planes on the top side with the ground plane on the backside of the flexible cable. The fast oscillations are due to standing waves within the ends of the cable, probably due to imperfect soldering of the SMA-connectors to the flexible part. Since we only use one half of the cable, as depicted in Fig. A\ref{appfig:S21}, the influence of the vias is significantly reduced and the oscillations are eliminated.

We use a microwave power of $\SI{14}{\dBm}$ for the 2D Rabi and Ramsey measurements and $\SI{10}{\dBm}$ for the CPMG measurements, where the power is measured directly in front of the cryostat.

In Fig. A\ref{appfig:S21} (b), the mixing chamber of the cryostat is depicted.  Microwave delivery occurs via the custom made flexible CPW, which is connected to the SC-CPW on the sample using aluminum bonding wires.

\subsection{\label{app:heating_effects}Heating effects of NC-CPW vs SC-CPW}
To qualitatively understand the heating effects of normalconducting (NC) versus superconducting (SC) coplanar waveguide antennas, we use the SnV center as a temperature sensor to roughly estimate the local temperature of the sample during microwave application. We note, that the measurements on NC-CPW are performed on SnV-A, which exhibits low strain, while the SC-CPW measurements are performed on the high strain SnV-B, that is used for the coherent control measurements in the main text.

\begin{figure}[b]
\includegraphics[width=\textwidth]{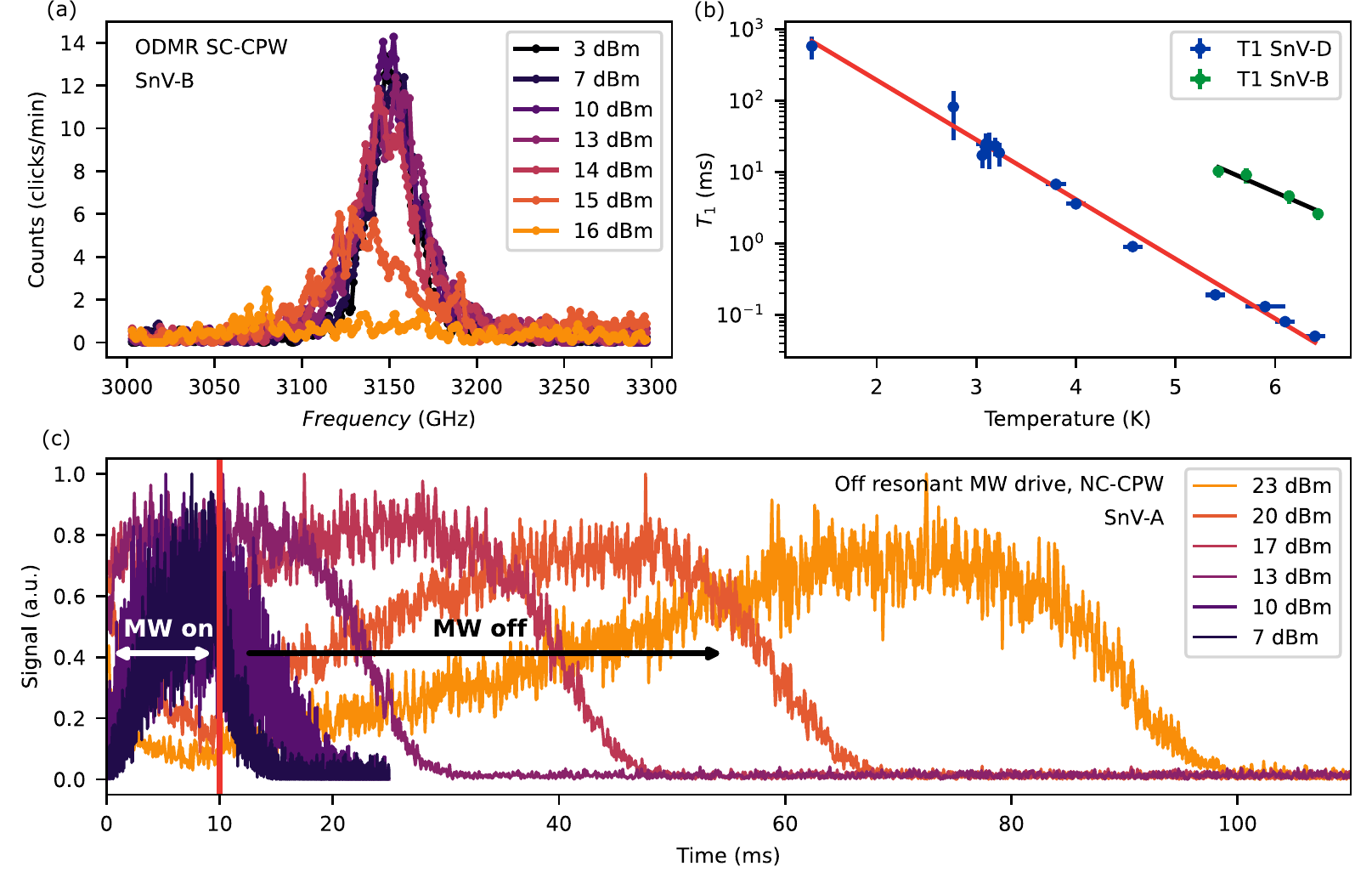}
\caption{\label{appfig:Heating_NC_CPW} Heating effect of NC-CPW versus SC-CPW and $T_1$ for an unstrained and a highly strained SnV center. (a) ODMR measurement of the highly strained SnV-B with the SC-CPW. The signal decays above $\SI{14}{\dBm}$, possibly due to the break of the superconductivity. The ODMR is measured by applying a repetitive MW chirp for $\SI{10}{\milli\second}$ followed by a $\SI{190}{\milli\second}$ waiting time. The total measurement duration is $\SI{2}{\min}$ for each ODMR. (b) $T_1$ measurement on an unstrained (blue points) and a highly strained (green points) SnV center. The strained SnV shows significantly higher spin lifetime due to the larger ground state splitting. (c) Heating effect of NC-CPW for off resonant cw-microwave drive at $\SI{3}{\giga\hertz}$ with $\SI{10}{\milli\second}$ duration followed by a $\SI{190}{\milli\second}$ break. The heightened local temperature causes increased countrate by thermal re-population of the continuously pumped readout state. Similar curves have been observed in \cite{Debroux2021}.}
\end{figure}

For SnV-A, we use a magnetic field of $\SI{193}{\milli\tesla}$ with an angle of $\theta = \SI{60}{\degree}$ to split the optically allowed transitions as well as the qubit transitions.
At this angle, the low strain SnV still exhibits high cyclicity and thus a high readout signal, but simultaneously the spin initialization time is reduced to the order of $\si{\micro\second}$ for excitation powers in the low $\si{\nano\watt}$ range.
Starting at a base temperature of $\SI{50}{\milli\kelvin}$, we apply an off-resonant microwave pulse with a length of $\SI{10}{\milli\second}$ and a frequency of $\SI{3}{\giga\hertz}$ to quantify the heating effect. We continuously initialize the SnV into the dark spin state by exciting the A1 transition. After the microwave pulse, we wait for $\SI{190}{\milli\second}$ to allow the heat to diffuse into the cryostat.
We repeat this sequence until the signal to noise ratio is satisfactory. We observe an increasing countrate during the microwave pulse, as is shown in Fig. A\ref{appfig:Heating_NC_CPW} (c).
Interestingly, for high microwave powers above $\SI{10}{\dBm}$, the countrate rises continuously even after the microwave is already switched off. To roughly estimate the local temperature of the SnV center, we compare to $T_1$ measurements previously performed on the unstrained SnV-D. At temperatures below $\SI{4}{\kelvin}$ the lifetime $T_1$ of the spin is already above one millisecond, as is shown in Fig. A\ref{appfig:Heating_NC_CPW} (b).
At the base temperature of the cryostat the spin lifetime is $T_1 > \SI{1}{\second}$. The increased countrate during the heating of the microwave line indicates, that the local temperature of the SnV center rises above $\SI{4}{\kelvin}$, causing the pumped state to be continuously thermally repopulated.

We attempted the same measurement for the high strain SnV-B with the SC-CPW, however no increase in countrate could be observed, even for microwave powers that break the superconductivity. We attribute this to the noticeably higher spin lifetime of SnV-B due to the larger ground state splitting. Fig. A\ref{appfig:Heating_NC_CPW} (b) illustrates the difference of the spin lifetimes for the unstrained and highly strained SnV centers, where SnV-B is showing a spin lifetime of already $\SI{2.59 \pm 0.45}{\milli\second}$ at $\SI{6.43 \pm 0.03}{\kelvin}$.

As an alternative, we perform ODMR measurements at high microwave powers, to understand the superconducting nature of the CPW. At a magnetic field of $\SI{93}{\milli\tesla}$ applied parallel to the quantization axis of SnV-B, we apply a microwave chirp for $\SI{10}{\milli\second}$ followed by a $\SI{190}{\milli\second}$ waiting time, similar to the measurement procedure described above. The resulting measurements are depicted in Fig. A\ref{appfig:Heating_NC_CPW} (a). We observe almost no change up to $\SI{14}{\dBm}$ microwave power, notice however, drastically decreasing signal for higher powers. The countrate does not increase during or after the microwave pulse, even after transitioning to the normalconducting state. Thus, we believe the temperature of the SnV center to remain below 6K. 

\clearpage

\section{\label{app:Hamiltonian}Electron spin Hamiltonian}
\subsection{Calibration of the DC magnetic field}
We initially estimate the magnetic field via COMSOL simulations, which typically exhibit low error. However, since the quenching factors are very sensitive to errors in the magnetic field strength, we further refine our measurements using the SnV as a sensor. To determine the magnetic field strength as precise as possible, we measure the response of the $^{13}\mathrm{C}$ nuclear spin-bath in both, parallel and perpendicular orientation to the SnV axis. This approach allows us to calculate the absolute current dependence of our magnetic field coils for all subsequent measurements. With a Larmor precession of $\gamma_{^{13}\mathrm{C}} = \SI{10.7084}{\mega\hertz\per\tesla}$ of the nuclear spin the strength of the applied magnetic field can be determined. Additionally, in the calibration process, we implement a field modulation technique, oscillating the magnetic field to asymptotically approach and stabilize at the target value. For the Spin-Echo measurement parallel to the SnV axis (Z-component), we fit to a single frequency coupling, arising from the $^{13}\mathrm{C}$ spin-bath:

\begin{equation}
    p_\parallel(\tau)=A \cdot \exp{\left(B\cdot \sin^4(2\pi f\cdot \tau+\varphi)-\frac{\tau}{T_\mathrm{damp}}\right)} + a\cdot \tau + \mathrm{DC}
\end{equation}

\begin{figure}
\includegraphics[width=\textwidth]{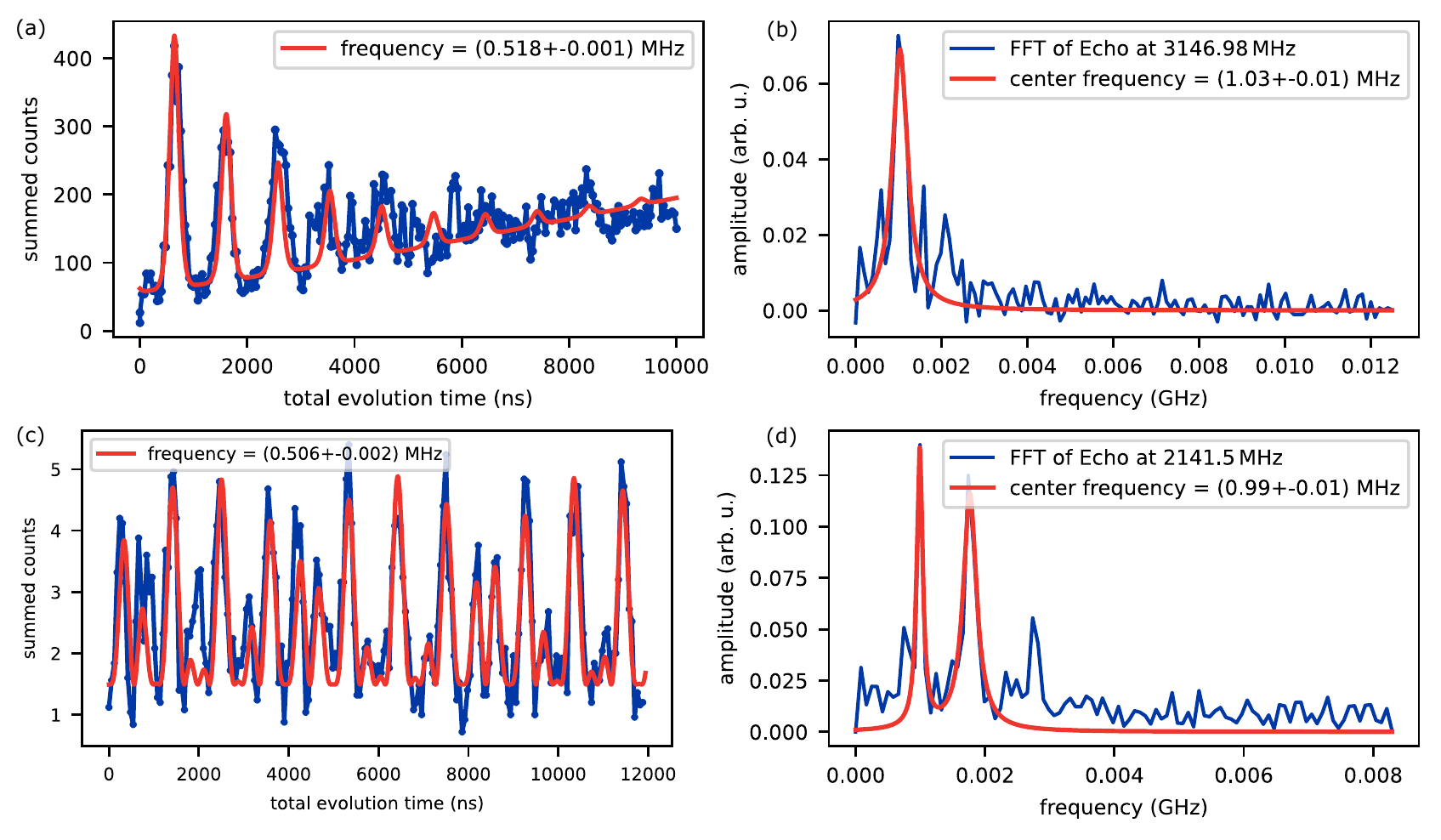}
\caption{\label{appfig:Echo_all} Hahn-Echo signal for parallel (a) and perpendicular (c) orientation of the magnetic field. The Echo is modulated by coupling to the ${}^{13}\mathrm{C}$ bath. For perpendicular orientation an additional coupling to a proximal spin is observed, resulting in a beat of the oscillations. The corresponding FFT signal is shown in (b) and (d), respectively.}
\end{figure}

The equation attributes phenomenological for additional coupling to other carbon-13 spins visible in the echo decay shown in figure A\ref{appfig:Echo_all} (a) by artificially introducing a damping time and a linear increase. The fit is in good agreement with a Lorentzian fit to the FFT of the echo signal, but yielding a better uncertainty compared to the FFT. For the Spin-Echo measurement perpendicular to the SnV axis ({\it y}-component), shown in A\ref{appfig:Echo_all} (c), we follow the same procedure. Interestingly, the coupling of the proximal nuclear spin is now visible, thus, we use the equation derived in \cite{childress2006} for NV centers coupled to a proximal $^{13}\mathrm{C}$ spin as a function:

\begin{equation}
   p_\perp(\tau) = A \cdot \sin\left(2\pi f_1\cdot\tau\right)^2 \cdot \sin\left(2\pi f_2 \cdot\tau\right)^2 + \mathrm{DC}
\end{equation}

One frequency corresponds to the spin-bath and one to the proximal nuclear spin. As reported in \cite{childress2006}, the precession of the proximal nuclear spin is enhanced, thus we attribute the higher frequency to it. The expected Larmor precession of the bath is given by

\begin{equation}
    f_\mathrm{larmor}=\frac{1}{2}\cdot \gamma_{^{13}\mathrm{C}} \cdot B=\frac{1}{2}\cdot\SI{10.7084}{\mega\hertz\per\tesla}\cdot \SI{100}{\milli\tesla}=\SI{0.53542}{\mega\hertz}
\end{equation}

The ratio between the measured frequency and the expected gives the corrected amplitude of the magnetic field 

\begin{align}
    B_\mathrm{par}&=\SI{96.7 \pm 0.2}{\milli\tesla}\, ,\\
    B_\mathrm{perp}&=\SI{94.5 \pm 0.4}{\milli\tesla}\, .
\end{align}

However, we overestimate the error for the magnetic field coils to be $0.5\%$ to account for effects like hysteresis for the following Hamiltonian fits and estimations of the orbital quenching factors. For the sake of completeness, we note the current dependency of our 3D-vector magnets in table \ref{apptab:3D_coils}, together with the vectors of the three SnVs (A-C) in the lab frame.

\begin{table}[b]
\caption{\label{apptab:3D_coils}
Estimated parameters of magnetic field coils spanning the lab frame and the SnV orientations within this lab frame.}
\begin{ruledtabular}
\begin{tabular}{cc|cc}
Coil axis & magnetic field ($\si{\milli\tesla\per\ampere}$) & SnV & quantization axis (X, Y, Z) \\
\hline
X & \SI{-45.7 \pm 0.2}{} & A & (-0.081, 0.834, -0.546)\\
Y & \SI{61.5 \pm 0.3}{}  & B & (0.011, 0.722, 0.692)\\
Z & \SI{133.5 \pm 0.6}{} & C & (-0.015, 0.883, -0.468)\\
   
\end{tabular}
\end{ruledtabular}
\end{table}

\clearpage
\subsection{\label{app:derivation_ham}Derivation of the Hamiltonian}
The Hamiltonian of this system is described by four contributions, namely the spin-orbit coupling, the orbital Zeeman-splitting, the Spin Zeeman-splitting and the strain splitting, leading to
\begin{equation}
\hat{H}_\mathrm{eff}^\mathrm{g,u} =  -\lambda^\mathrm{g,u} \hat{L}_z \hat{S}_z + \mu_B \left(f_{m_j}^\mathrm{g,u}\right)_{\{\pm\frac{1}{2},\pm\frac{3}{2}\}}\cdot \hat{L}_z B_z  + g_S \mu_B \bm{\hat{S}\cdot \hat{B}} + \hat{\Upsilon}_\mathrm{strain}
\end{equation}
The orbital quenching factor is defined as $f_{m_j}^\mathrm{g,u}=g_L^\mathrm{g,u}\cdot p_{m_j}^\mathrm{g,u}$ introduced in \cite{Thiering_magneto_optic}, where $g_L^{g,u}$ is the Stevens reduction factor for ground and excited state and $p_{m_j}$ the Ham-reduction factor for each effective total angular momentum quantum number $m_j=\{\pm\frac{1}{2}, \pm\frac{3}{2}\}$. As we are not able to differentiate these two components in our measurements, we stick to the common orbital quenching factor $f_{m_j}^\mathrm{g,u}$. To derive a matrix representation of the Hamiltonian, for simplicity, we show the derivation for the ground state, which applies equivalently to the excited state. The $\hat{L}_z$ operator in the $\{xy\}$-basis of the electronic states, where the $m_l=0$ $(A_{2u})$ orbital is far detuned and thus irrelevant \cite{Hepp_2014}, is given by the matrix 
\begin{equation}
    \hat{L}_z=\begin{pmatrix}
        0 & i \\
        -i & 0
    \end{pmatrix}\,.
\end{equation}
The spin-orbit splitting Hamiltonian follows as
\begin{equation}
    H_\mathrm{so}^{xy}=-\lambda^\mathrm{g} \hat{L}_z \hat{S}_z =\left[\begin{matrix}0 & 0 & - \frac{i \lambda^\mathrm{g}}{2} & 0\\0 & 0 & 0 & \frac{i \lambda^\mathrm{g}}{2}\\\frac{i \lambda^\mathrm{g}}{2} & 0 & 0 & 0\\0 & - \frac{i \lambda^\mathrm{g}}{2} & 0 & 0\end{matrix}\right]\, ,
\end{equation}
by performing the outer product between the $\hat{L}_z$ operator and the $\sigma_z$ Pauli-matrix. 

The transformation matrix $T$, which diagonalizes $H_\mathrm{so}^{xy}$ and thus connects the $\{xy\}$-basis and spin-orbit eigenbasis or short $\{so\}$-basis, is given by the relation
\begin{equation}
    T^{-1}\cdot H_\mathrm{so}^{xy}\cdot T = H_\mathrm{so}^{so}\, .
\end{equation}
Solving this gives equation results in the matrix 
\begin{equation}
    T=\left[\begin{matrix}i & 0 & - i & 0\\0 & - i & 0 & i\\1 & 0 & 1 & 0\\0 & 1 & 0 & 1\end{matrix}\right]\, ,
\end{equation}
and the diagonalized spin-orbit coupling follows as
\begin{equation}
    H_\mathrm{so}^{so} = \left[\begin{matrix}- \frac{\lambda^\mathrm{g}}{2} & 0 & 0 & 0\\0 & - \frac{\lambda^\mathrm{g}}{2} & 0 & 0\\0 & 0 & \frac{\lambda^\mathrm{g}}{2} & 0\\0 & 0 & 0 & \frac{\lambda^\mathrm{g}}{2}\end{matrix}\right]\,.
\end{equation}
The Spin Zeeman-term without any asymmetry contributes to the Hamiltonian as
\begin{equation}
\bar{H}_\mathrm{Ze}^{xy} = \frac{\gamma_s}{2}\left[\begin{matrix}B_\parallel & B_\perp & 0 & 0\\B_\perp & - B_\parallel  & 0 & 0\\0 & 0 & B_\parallel & B_\perp \\0 & 0 & B_\perp & - B_\parallel \end{matrix}\right]\,,
\end{equation}
where the parallel magnetic field orientation is chosen along the SnV center quantization axis.

For the orbital Zeeman term one has to account for the orbital quenching factors $f_{m_j}$ for each sub-level with their distinct orbital quantum number $m_j=\{\pm\frac{1}{2}, \pm\frac{3}{2}\}$. For the ordering of the sub-states, we follow \cite{Thiering_magneto_optic}, which leads in the spin-orbit Eigenbasis to the matrix
\begin{equation}
   H_\mathrm{Ze}^{so} = \gamma_l B_\parallel\left[\begin{matrix}f_{32} & 0 & 0 & 0\\0 & - f_{32}  & 0 & 0\\0 & 0 & -f_{12}  & 0\\0 & 0 & 0 & f_{12}\end{matrix}\right]\, ,
\end{equation}
where we omitted the fraction in the subscript for better readability.
Transforming this into the $\{xy\}$-basis using the transformation matrix $T$, results in
\begin{equation}
\small
  H_\mathrm{ZL}^{xy} = \frac{\gamma_l B_\parallel}{2}\left[\begin{matrix}- f_{12}  +f_{32}  & 0 & i f_{12}  + i  f_{32}  & 0\\0 & f_{12}  - f_{32}  & 0 & i  f_{12} + i f_{32} \\- i  f_{12}  - i  f_{32} & 0 & -  f_{12}  +  f_{32} & 0\\0 & - i  f_{12}- i  f_{32}  & 0 &  f_{12}  -  f_{32}\end{matrix}\right]\,.
\end{equation}
This result is already hinting the emerging of mean values of the reduction factors $f$ and differences leading to finite asymmetries $\delta$, like reported in \cite{Thiering_magneto_optic}. For the full Hamiltonian, the strain contribution in the $\{xy\}$-basis has to be accounted for, which is given by
\begin{equation}
    H_\mathrm{JT}^{xy}=\left[\begin{matrix}\alpha & 0 & 0 & 0\\0 & \alpha & 0 & 0\\0 & 0 & - \alpha & 0\\0 & 0 & 0 & - \alpha\end{matrix}\right]\, ,
\end{equation}
including strain and Jahn-Teller effects in the coupling constant $\alpha$ (see e.g. \cite{Hepp_2014}). Collecting everything gives the full electronic Hamiltonian in the $\{xy\}$-basis as
\begin{equation}
\small
H_{xy} = \frac{1}{2}
\left[\begin{matrix} (\gamma_l(f_{32} - f_{12}) + \gamma_{s}  )B_\parallel + 2 \alpha & \gamma_s B_\perp & i\gamma_l (f_{12} + f_{32} )B_\parallel  - i \lambda^\mathrm{g} & 0\\

\gamma_s B_\perp & (\gamma_{l} (f_{12} - f_{32}) - \gamma_{s})B_\parallel  + 2 \alpha & 0 & i\gamma_{l} (f_{12}  +f_{32})B_\parallel + i \lambda^\mathrm{g}\\

(- i \gamma_{l} (f_{12}+ f_{32}))B_\parallel  + i \lambda^\mathrm{g} & 0 & ( \gamma_{l}(f_{32} - f_{12}) + \gamma_{s} )B_\parallel - 2 \alpha & \gamma_s B_\perp

\\0 & - i\gamma_{l}( f_{12} + f_{32})B_\parallel - i \lambda^\mathrm{g} & \gamma_s B_\perp & (\gamma_{l}(f_{12}  - f_{32}) - \gamma_{s} )B_\parallel - 2 \alpha\end{matrix}\right]\, .
\end{equation}
The Eigenvalues of this matrix are used to fit the measured ODMR data, the difference in the allowed transitions $A1$ and $B2$ and, if measured, the difference in the forbidden transitions $B1$ and $A2$. For the sake of completeness the Hamiltonian in the $\{so\}$-basis is given by
\begin{equation}
\small
H_\mathrm{so} = \frac{1}{2}
\left[\begin{matrix} (2 f_{32} \gamma_{l} + \gamma_{s} ) B_\parallel - \lambda^\mathrm{g} & 0 & - 2 \alpha & \gamma_{s} B_\perp \\0 & (- 2  f_{32} \gamma_{l} -  \gamma_{s} )B_\parallel - \lambda^\mathrm{g} &  \gamma_{s} B_\perp & - 2 \alpha\\- 2 \alpha & \gamma_{s}B_\perp  & (- 2 f_{12} \gamma_{l} + \gamma_{s} ) B_\parallel + \lambda^\mathrm{g} & 0\\\gamma_{s} B_\perp  & - 2 \alpha & 0 & (2  f_{12} \gamma_{l} - \gamma_{s} )B_\parallel + \lambda^\mathrm{g}\end{matrix}\right]\, .
\end{equation}
We note that we can rewrite this Hamiltonian by introducing the commonly used mean reduction factors $f = \frac{f_{12}}{2}+\frac{f_{32}}{2}$, and the mean asymmetries $\delta = \frac{f_{32}}{2}-\frac{f_{12}}{2}$, and $\gamma_l = \frac{\gamma_s}{2}$, leading to
\begin{equation}
H_\mathrm{so} = 
\left[\begin{matrix} (f\gamma_{l}+\delta\gamma_{l} + \frac{\gamma_{s}}{2} ) B_\parallel - \frac{\lambda^\mathrm{g}}{2} & 0 & -  \alpha & \frac{\gamma_{s} B_\perp}{2} \\0 & (- f\gamma_{l} - \delta\gamma_{l} -  \frac{\gamma_{s}}{2} )B_\parallel - \frac{\lambda^\mathrm{g}}{2} &  \frac{\gamma_{s} B_\perp}{2} & -  \alpha\\- \alpha & \frac{\gamma_{s}B_\perp}{2}  & (- f\gamma_{l} +\delta\gamma_{l} + \frac{\gamma_{s}}{2} ) B_\parallel + \frac{\lambda^\mathrm{g}}{2} & 0\\\frac{\gamma_{s} B_\perp}{2}  & -  \alpha & 0 & (f\gamma_{l}-\delta\gamma_{l} - \gamma_{s}/2 )B_\parallel +\frac{\lambda^\mathrm{g}}{2}\end{matrix}\right]\,.
\end{equation}
This matrix is, by accounting the freedom to rearrange the columns, the same Hamiltonian as reported in \cite{Thiering_magneto_optic}.

\subsection{Fitting procedure and uncertainty estimation}
\label{appsub:Fitting_procedure}
In our procedure the Eigenvalues of the full Hamiltonian are numerically fitted. However, for an intuitive understanding of the different parameters, we analytically calculate the Eigenvalues for different strain and magnetic field regimes. First, we look at the qubit transitions for negligible strain $\alpha^\mathrm{g}$, in comparison to the spin-orbit coupling $\lambda^\mathrm{g}$. This gives, for parallel orientation of the magnetic field, an ODMR splitting of
\begin{equation}
    \ket{2}-\ket{1} = B_\parallel\left(2f_{32}^\mathrm{g}\gamma_l + \gamma_s\right)\, ,
\end{equation}
showing that the qubit transitions at parallel field only depend the orbital quenching factor $f_{32}^\mathrm{g}$ in the case of vanishing strain. Further, under the same conditions, one finds for the allowed transitions
\begin{equation}
    A1-B2=2B_\parallel(f_{32}^\mathrm{g}-f_{32}^\mathrm{u})\gamma_l\,.
\end{equation}
Hence, by using the value of $f_{32}^\mathrm{g}$, that we obtain from fitting the qubit transitions, the excited state quenching factor $f_{32}^\mathrm{u}$, can be determined by fitting the allowed transitions.

To determine the exact strain magnitude, one can look at the splitting of the qubit at perpendicular field orientation, which follow as
\begin{equation}
    \ket{2}-\ket{1} =\frac{1}{2}\left(\sqrt{(\gamma_s B_\perp-2\alpha^\mathrm{g})^2+(\lambda^\mathrm{g})^2}-\sqrt{(\gamma_s B_\perp+2\alpha^\mathrm{g})^2+(\lambda^\mathrm{g})^2}\right)\, .
\end{equation}
The excited state strain can be found equivalently, via the allowed transitions
\begin{equation}
\small
    A1-B2  = \frac{1}{2} \left(\sqrt{(\gamma_s B_\perp-2\alpha^\mathrm{u})^2+(\lambda^\mathrm{u})^2}-\sqrt{(\gamma_s B_\perp+2\alpha^\mathrm{u})^2+(\lambda^\mathrm{u})^2}-\sqrt{(\gamma_s B_\perp-2\alpha^\mathrm{g})^2+(\lambda^\mathrm{g})^2}+\sqrt{(\gamma_s B_\perp+2\alpha^\mathrm{g})^2+(\lambda^\mathrm{g})^2}\right)\, .
\end{equation}
For higher strain magnitude, we can determine the quenching factors $f_{12}^\mathrm{g,u}$, as the Eigenvalues now also depend on this value. For the qubit transition, we find
\begin{equation}
    \small
    \ket{2}-\ket{1} =\frac{1}{2}\left(\sqrt{(\lambda^\mathrm{g}-\gamma_l B_\parallel(f_{12}^\mathrm{g}+f_{32}^\mathrm{g})^2)+4(\alpha^\mathrm{g})^2}-\sqrt{(\lambda^\mathrm{g}+\gamma_l B_\parallel(f_{12}^\mathrm{g}+f_{32}^\mathrm{g})^2)+4(\alpha^\mathrm{g})^2}
    + 2\gamma_l B_\parallel (f_{12}^\mathrm{g}-f_{32}^\mathrm{g})-2B_\parallel\gamma_s\right)\,,
\end{equation}
which allows us to determine $f_{12}^\mathrm{g}$.
With the same approach as before, we can now determine the last quenching factor $f_{12}^\mathrm{u}$ via the allowed transitions.
We omit the analytical expression, due to the size of the equation.

\begin{figure}[b]
\includegraphics[width=\textwidth]{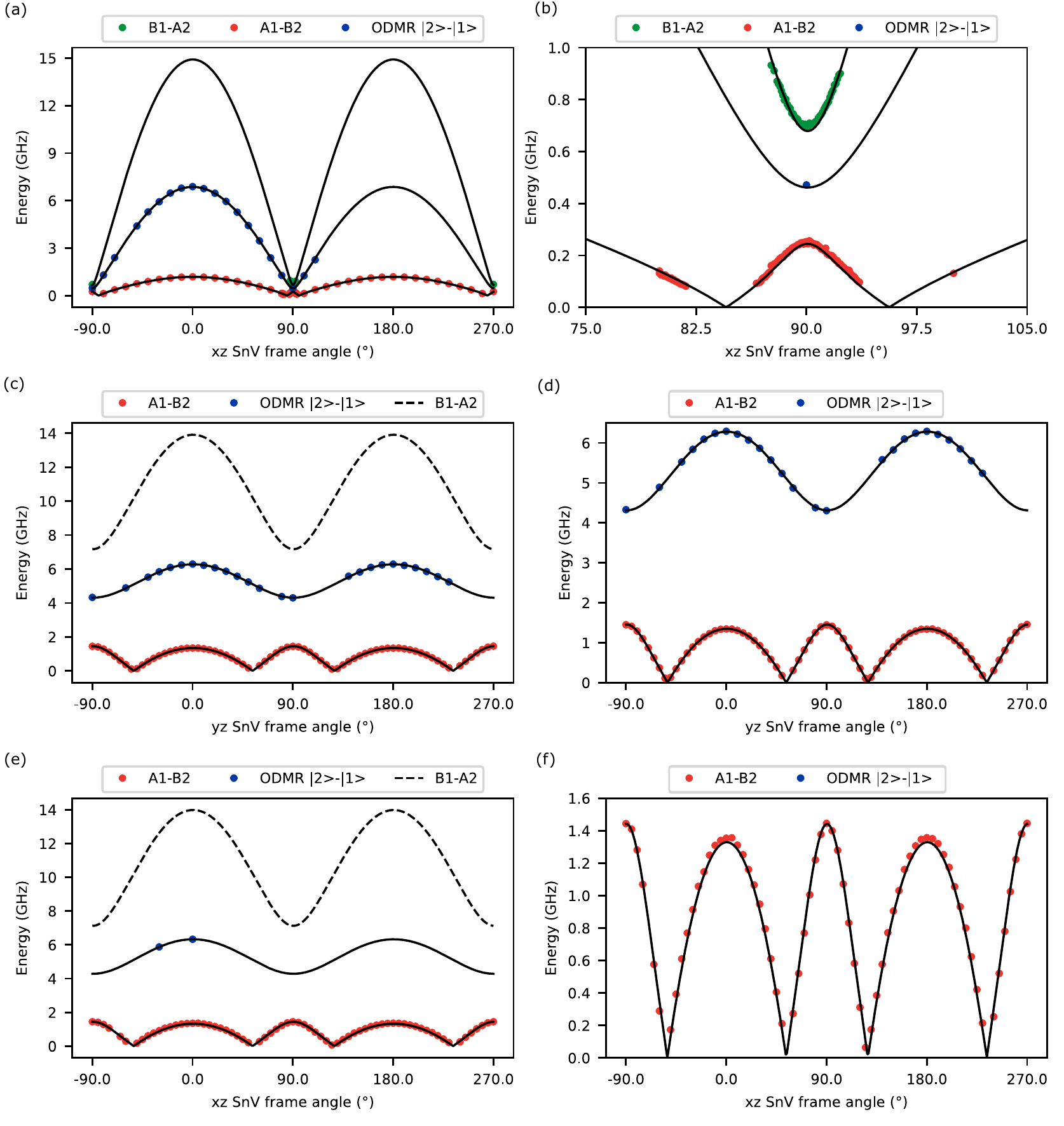}
\caption{\label{appfig:Ham_fit_all} Fit of the Hamiltonian. (a-b) Low strain to SnV A. This emitter is used to determine $f_{\frac{3}{2}}^\mathrm{g,u}$. Due to the low branching ratio at perpendicular magnetic field orientation ($\theta = \SI{90}{degree}$), the forbidden transitions B1-A2 are detectable (black dots) at temperatures where the spin lifetime is much shorter that the sweeping speed of the PLE measurement. (c-d) high strain SnV-B. This emitter is used to determine $f_{\frac{1}{2}}^\mathrm{g,u}$. (c-d) high strain SnV-C. This emitter is used to validate the fitting parameters $f_{\frac{1}{2}, \frac{3}{2}}^\mathrm{g,u}$.}
\end{figure}

\begin{figure}[b]
    \centering
    \includegraphics[width=\textwidth]{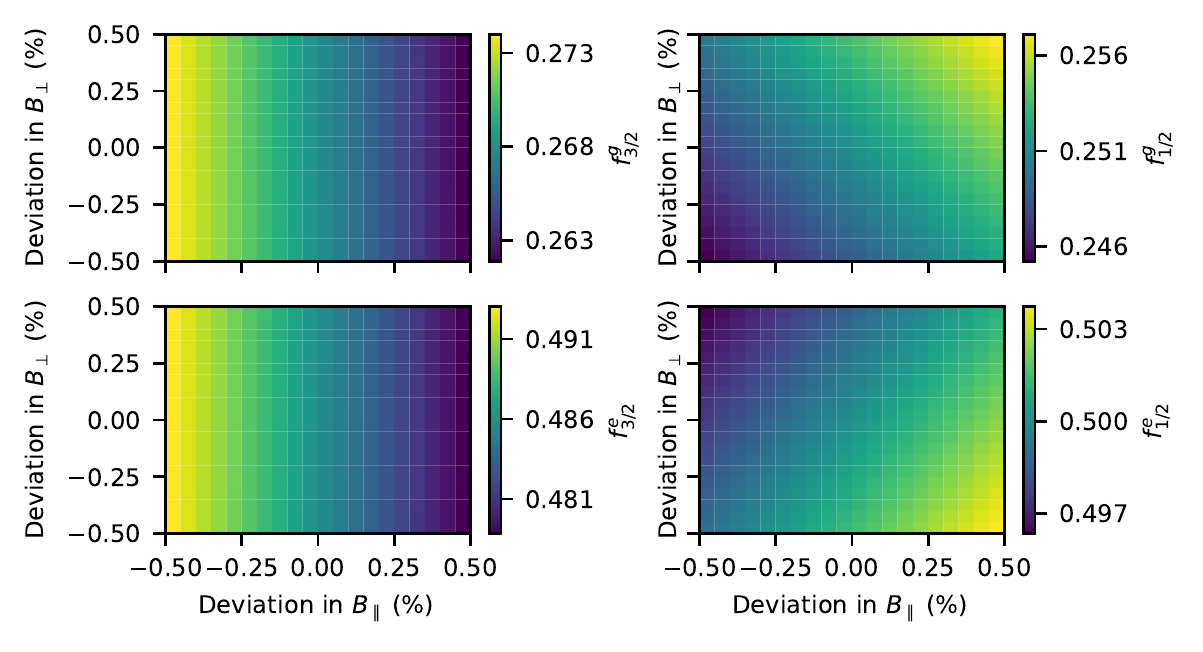}
    \caption{Spread of the quenching $f$ parameters from fit to SnV-A and SnV-B, assuming an uncertainty of $\pm 0.5\%$ on the $B$-field amplitude. For each magnetic field strength the Hamiltonian is fit and the resulting $f$ parameter plotted in the corresponding pixel. As discussed in section \ref{app:Hamiltonian}, first $f_{3/2}^\mathrm{g}$ is fitted, followed by $f_{1/2}^\mathrm{g}$, $f_{3/2}^\mathrm{e}$, and $f_{1/2}^\mathrm{e}$, each taking the previous quenching parameter into account. The total spread of the $f$ parameters in each subplot is taken as the standard uncertainty for this $f$ parameter.}
    \label{appfig:f-errors}
\end{figure}

The resulting fits of the three different investigated SnV centers to the Hamiltonian Eigenvalues are shown in Fig. A\ref{appfig:Ham_fit_all}. With the estimated error on the magnetic field magnitude of $0.5\%$, we evaluate the resulting quenching factors with a varying amplitude of $B_{\mathrm{dc}}$ given by
\begin{equation}
    B^{\parallel, \perp}_\mathrm{DC}(x) = B^{\parallel, \perp}_\mathrm{mean}\cdot (1 + x \cdot 0.5\%),\quad x \in \{-1,1\}\, .
\end{equation}
We omit errors on the fit of the allowed transitions, as well as on the ODMR transitions as these uncertainties are at least an order of magnitude smaller than the obtained values and are negligible compared to the uncertainty stemming from the magnetic field amplitude. As the influence of the error on $\lambda^g$ and thus on the strain $\alpha^g$ is also small compared to the uncertainty on the magnetic field magnitude, this uncertainty is also not propagated onto the quenching parameters. The result of the fitting procedure is shown in Fig. A\ref{appfig:f-errors}. The range of the obtained quenching parameters is taken as the standard deviation for our estimation.

\clearpage
\subsection{Optical Properties of the SnV center}
\begin{figure}[b]
\includegraphics[width=\linewidth]{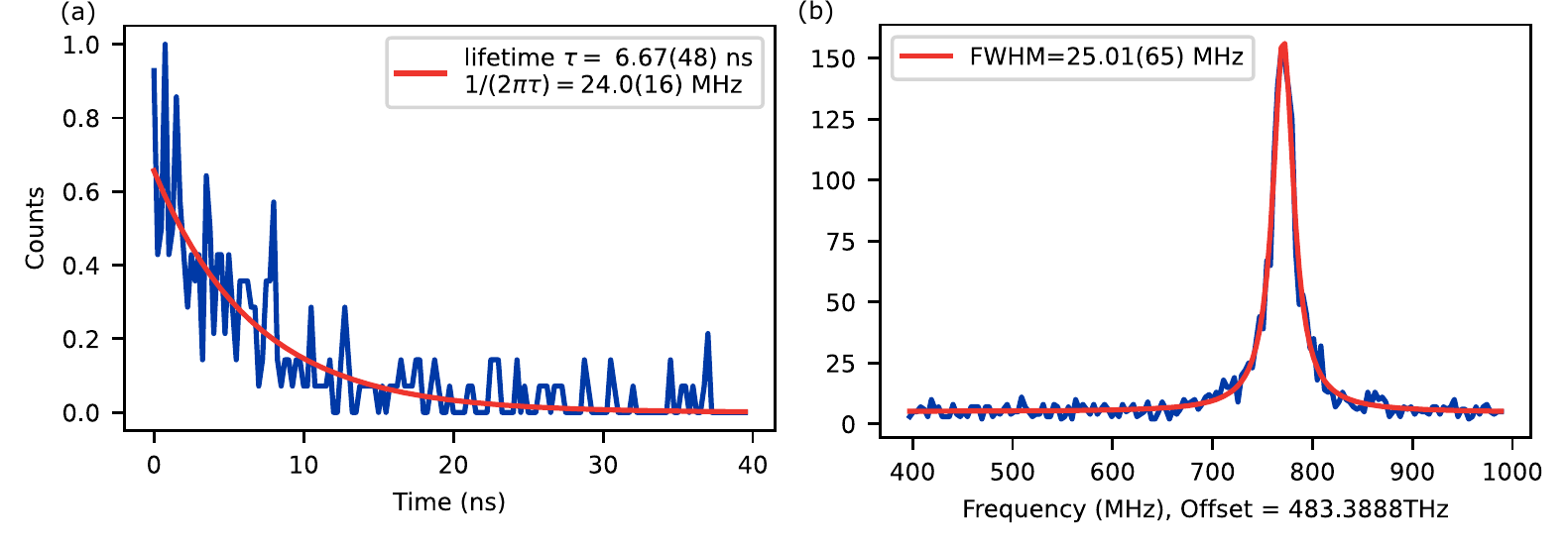}
\caption{\label{appfig:FHWM} Single scan PLE linewidth (left) and pulsed resonant lifetime measurement (right). Optical linewidth is given by the lifetime and thus Fourier-limited.}
\end{figure}

\begin{figure}[b]
\includegraphics[width=\textwidth]{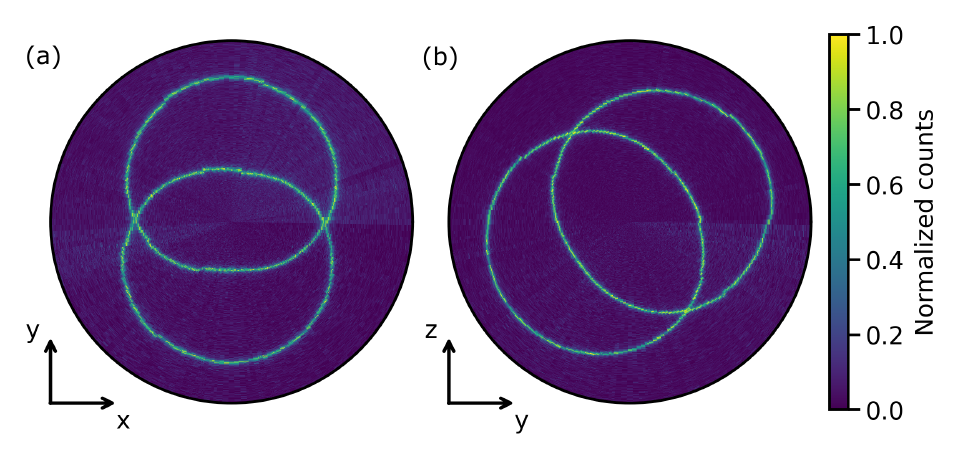}
\caption{\label{appfig:Polar_SnV_d} Polar plots of the unstrained SnV-D in the lab frame. (a) Splitting of the A1 and B2 transitions in the \textit{xy} axis of the laboratory frame of SnV-D. The PLE scans are shown in polar coordinates, where the radial axis spans a \SI{1.5}{\giga\hertz} range. (b) Measurement in the \textit{yz} axis and a PLE span of \SI{2}{\giga\hertz}. In contrast to strained SnV centers, no splitting is visible for perpendicular field orientation.}
\end{figure}

In Fig. A\ref{appfig:FHWM} (a), we show the lifetime measurement of SnV-B along with a single PLE scan, demonstrating the excellent optical properties of the SnV center even under the influence of strain.
We measure a lifetime of $\SI{6.67 \pm 0.48}{\nano\second}$, in accordance to the transform limited linewidth in Fig. A\ref{appfig:FHWM} (b). The polar plots of the unstrained emitter SnV-D are depicted in Fig. A\ref{appfig:Polar_SnV_d} (a,b) in the \textit{xy} and \textit{yz} lab frame. In contrast to strained SnV centers, the degeneracy of the allowed transition A1 and B2 is not lifted, for perpendicular field orientation.

\clearpage
\section{\label{app:Coh_contr}Coherent Control}
We perform the coherence measurements in a parallel magnetic field of \SI{93}{\milli\tesla} and applying a resonant microwave with \SI{10}{dBm} at \SI{3149.1}{\mega\hertz}, resulting in a  \SI{196}{\nano\second} long $\pi$-pulse. Reading out and initializing involve a resonant pulse of $\qtyrange{1.2}{2.0}{\milli\second}$. After each data point, the charge state is assessed through threshold counting and a green repump pulse is applied, if necessary. However, no green re-pump is applied during the dynamical decoupling sequence. We apply a $+\frac{\pi}{2}$-pulse at the end of the decoupling sequence in-phase relative to the initial $+\frac{\pi}{2}$-pulse. The recorded data is the cumulative count over the entire readout time. Further, we normalize the data such, that the final saturating points (i.e. last 10 data points taken) are corresponding to a Fidelity of \num{0.5} \cite{SukachevSiV}. We fit the normalized data to the function
\begin{equation}
    A \cdot \exp\left(-\left(\frac{N\tau}{T_2}\right)^\mathrm{\xi}\right) + 0.5\, ,
\end{equation}
where $A$ represents the amplitude, $N$ are the number of decoupling pulses, and $\xi$ denotes the stretching factor, set at a constant value of $4$ to achieve optimal agreement with the data. The resulting decay curves are shown in Fig. \ref{fig:fig_cpmg}. Plotting the obtained coherence times over the number of decoupling pulses, we obtain a exponential increase in $T_2$ as shown in Fig. \ref{fig:fig_cpmg}.

\begin{figure}[b]
\includegraphics[width=\linewidth]{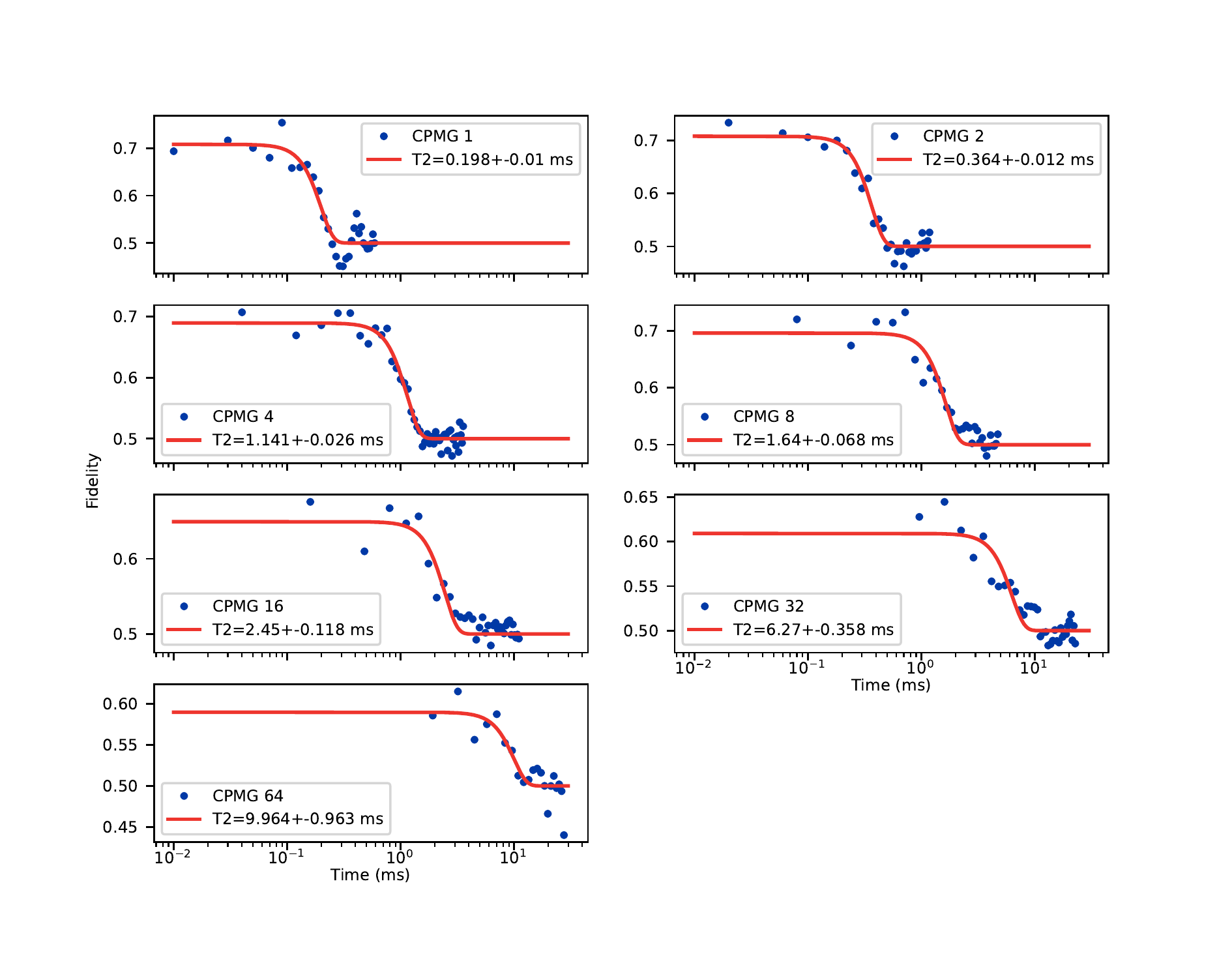}
\caption{\label{appfig:fig_cpmg_all_collection} CPMG sequences with varying number $N$ refocusing pulses. The data is scaled to \num{0.5} for long times and fit to stretched exponential envelopes with $e^{-(t/T_2)^{4}}$.}
\end{figure}

\clearpage

\section{\label{app:strain}Adhesive-induced strain analysis}
\begin{figure}[b]
\includegraphics[width=\textwidth]{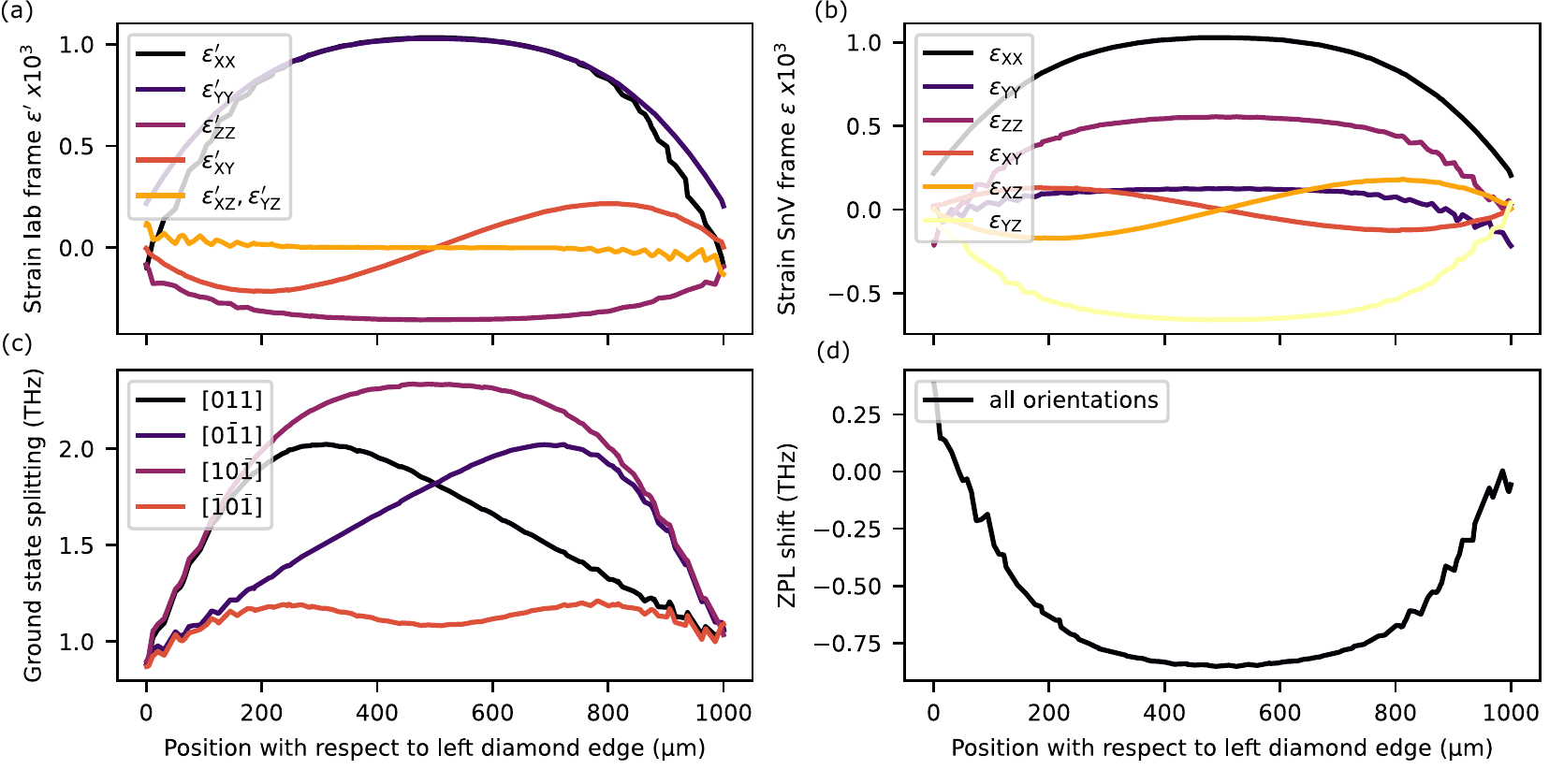}
\caption{\label{appfig:app_strain_comsol} Strain simulation results. (a) Strain magnitudes for different cartesian directions in the lab frame. (b) Rotated strain magnitudes with respect to the exemplary possible SnV axis $[011]$. (c) Calculated ground state splitting for all four SnV center orientations. We denote the orientations of the SnV center axis within a diamond with $\left<110\right>$ edges, corresponding to the set $[111,\ 1\bar{1}\bar{1},\ \bar{1}1\bar{1},\ \bar{1}\bar{1}1]$. (d) Calculated shift of the ZPL as a function of the position of the SnV center along the cut shown in Fig. \ref{fig:fig_overwiev} (f). Strain susceptibilities of the SiV (see reference \cite{Meesala_strain}) are used to calculate the shift. Due to to symmetry all orientations observe the same splitting.}
\end{figure}

To evaluate the strain magnitude induced by the optical adhesive in the diamond, we use the software COMSOL. We assume a diamond thickness of \SI{26}{\micro\meter} and a polymer thickness of \SI{150}{\micro\meter} spreading a diameter of \SI{2}{\milli\meter}. The diamond is assumend to be emerged \SI{5}{\micro\meter} deep into the polymer. Since the coefficient of thermal expansion (CTE) for NOA63 is unknown, we use the CTE from the similar adhesive NOA61 \cite{noa61}. For simplicity, we approximate the CTE of the adhesive by a step-like function with expansion coefficients $\left[230,\ 90,\  80\right] \times 10^{-6} \si{\per\kelvin}$ changing at temperature steps $\left[270,\ 230,\ 4\right] \si{\kelvin}$ and a linear transition range over $\frac{1}{5}$ of the temperature range between each step. The modulus of elasticity is taken constant as \SI{1.6}{\giga\pascal} for the whole temperature range, given as a typical value by the manufacturer \cite{noa63}. The resulting strain components are calculated over the whole diamond surface. The results are shown in Fig. A\ref{appfig:app_strain_comsol}. To convert the strain components $\epsilon^\prime$ into the reference frame of the SnV center, the tensor is rotated according to
\begin{equation}
    \epsilon = R_z(\theta) R_y(\phi)\ \epsilon^\prime\ R_y(\phi)^\dagger R_z(\theta)^\dagger\, ,
\end{equation}
with the angles $\theta=\SI{90}{\degree}$ and $\phi=\SI{54}{\degree}$. Due to the different orientations, we find differences in the resulting values for all $\left<111\right>$ directions. Using the definitions of the strain components with respect to the symmetry
\begin{align}
    \epsilon_{A_1} & = t_\perp \left(\epsilon_{xx} + \epsilon_{yy}\right) + t_\parallel \epsilon_{zz}\\
    \epsilon_{E_x} & = d (\epsilon_{xx} - \epsilon_{yy}) + f \epsilon_{zx}\\
    \epsilon_{E_y} & = -2d \epsilon_{xy} - f \epsilon_{yz}\, ,
\end{align}
and the susceptibilities $f=\SI{-0.56e6}{\giga\hertz}$ and $d=\SI{0.8e6}{\giga\hertz}$ \cite{Guo} the strain tensor
\begin{equation}
    H_\mathrm{strain}=\begin{pmatrix} \epsilon_{A1} - \epsilon_{E_x} & \epsilon_{E_y} \\ \epsilon_{E_y} & \epsilon_{A1} + \epsilon_{E_x} \end{pmatrix} \otimes \begin{pmatrix} 1 & 0 \\ 0 & 1 \end{pmatrix}\, ,
\end{equation}
can be diagonalized. Finally, the corresponding ground state splittings are calculated, by taking the differences of the Eigenvalues. To calculate the shift of the ZPL line, we use the measured values of the SiV $t_\parallel = \SI{-1.7e6}{\giga\hertz}$ and $t_\perp= \SI{0.078e6}{\giga\hertz}$ taken from \cite{Meesala_strain}. The shift of the ZPL  is shown in Fig. A\ref{appfig:app_strain_comsol} and in qualitative agreement with the \SI{1}{\nano\meter} shift observed over the half of the diamond shown in Fig. \ref{fig:fig_overwiev}. As the simulated ground state splitting shows a more complex substructure than the measured ground state splitting, we have to emphasize that this simulation is just a qualitative analysis. Better results can be obtained by using a specified geometry \cite{Guo} or simulating with measured and not assumend properties of the polymer.

\clearpage

\end{document}